\documentclass[conference]{IEEEtran}
\IEEEoverridecommandlockouts

\usepackage[cmex10]{amsmath}
\usepackage{amssymb}
\usepackage{graphicx}
\usepackage{subfigure}
\usepackage{cite}
\usepackage{epstopdf}
\usepackage{subeqnarray}
\usepackage{diagbox}
\usepackage{booktabs}
\usepackage{multirow}
\usepackage{multicol}
\usepackage{algorithm}
\usepackage{algorithmic}
\usepackage{setspace}
\usepackage{url}
\usepackage{xcolor}
\usepackage{array}
\usepackage{footnote}
\usepackage{enumerate}
\usepackage{framed}
\usepackage{lineno}
\usepackage{threeparttable}
\makesavenoteenv{tabular}

\newcommand{\mb}{\mathbf}
\newcommand{\mbb}{\mathbb}
\newcommand{\mc}{\mathcal}

\newcommand{\mr}{\mathrm}
\newcommand{\ti}{\textit}

\newtheorem{theorem}{\textbf{Theorem}}

\newtheorem{proposition}{\textbf{Proposition}}
\newtheorem{corollary}{\textbf{Corollary}}

\newtheorem{example}{\textbf{Example}}

\begin{document}

\title{Mutual Information-Maximizing Quantized Belief Propagation Decoding of Regular LDPC Codes}

\author{%
\IEEEauthorblockN{Xuan He$^\dag$, Kui Cai$^\ddag$, Zhen Mei$^*$, Peng Kang$^\ddag$, and Xiaohu Tang$^\dag$}
\IEEEauthorblockA{$^\dag$Southwest Jiaotong University, Chengdu {\rm 611756}, China\\
$^\ddag$Singapore University of Technology and Design, Singapore {\rm 487372}, Singapore\\
$^*$Nanjing University of Science and Technology, Nanjing {\rm 210094}, China\\
Email: xhe@swjtu.edu.cn, cai\_kui@sutd.edu.sg,  meizhen@njust.edu.cn, peng\_kang@sutd.edu.sg, xhutang@swjtu.edu.cn}
}

\maketitle

\begin{abstract}
In this paper, we propose a class of finite alphabet iterative decoder (FAID), called mutual information-maximizing quantized belief propagation (MIM-QBP) decoder, for decoding regular low-density parity-check (LDPC) codes. Our decoder follows the reconstruction-calculation-quantization (RCQ) decoding architecture that is widely used in FAIDs. We present the first complete and systematic design framework for the RCQ parameters, and prove that our design with sufficient precision at node update is able to maximize the mutual information between coded bits and exchanged messages. Simulation results show that the MIM-QBP decoder can always considerably outperform the state-of-the-art mutual information-maximizing FAIDs that adopt two-input single-output lookup tables for decoding. Furthermore, with only 3 bits being used for each exchanged message, the MIM-QBP decoder can outperform the floating-point belief propagation  decoder at the high signal-to-noise ratio regions when testing on high-rate LDPC codes with a maximum of 10 and 30 iterations.
\end{abstract}


\IEEEpeerreviewmaketitle


\section{Introduction}
Low-density parity-check (LDPC) codes\cite{Gallager62} have been widely applied to communication and data storage systems due to their capacity approaching performance.
Many of these systems, such as the NAND flash memory, have strict requirements on the memory consumption and implementation complexity of LDPC decoders \cite{chen2018rate, aslam2017edge}.
For the sake of simple hardware implementation, many efforts have been devoted to efficiently represent messages for LDPC decoding
\cite{Chen05, Zhang2009qbp, Richardson01capacity, Lee05, Thorpe02, Romero16, Lewandowsky18, meidlinger2020design}.
For example, in \cite{Chen05} and \cite{Zhang2009qbp}, the messages exchanged within the LDPC decoders are represented by log-likelihood ratios (LLRs) in low resolutions, which have generally 5 to 7 bits.
The works in \cite{Richardson01capacity, Lee05, Thorpe02, Romero16, Lewandowsky18, meidlinger2020design} develop the finite alphabet iterative decoder (FAID) for LDPC codes, which make use of messages represented by symbols from finite alphabets rather than by LLRs for decoding.
These FAIDs draw much attention due to their excellent error rate performance and low decoding complexity.
More specifically, they adopt the quantized messages of only 3 or 4 bits  to approach and even surpass the performance of the floating-point belief propagation (BP) decoder \cite{MacKay1999SPA}.
In this paper, we focus on the class of FAIDs.

The early FAIDs investigated in \cite{Richardson01capacity, Lee05, Thorpe02} consider simple mappings, additions, and non-uniform quantization for decoding, leading to a reconstruction-calculation-quantization (RCQ) decoding architecture.
However, the design of these FAIDs (RCQ parameters) focuses on the specific class of (3, 6) (variable node (VN) degree 3 and check node (CN) degree 6) LDPC codes  and requires large amount of manual optimizations.
Thus, one can hardly generalize the decoder design to different scenarios.
Recently, the mutual information-maximizing FAIDs (MIM-FAIDs) \cite{Romero16, Lewandowsky18, meidlinger2020design} are designed by density evolution (DE) \cite{Richardson01capacity} for various bit width settings of the decoder, with the aim of maximizing  the mutual information (MI) between the coded bits and the exchanged messages within the decoder.
These MIM-FAIDs are also referred to as MIM lookup table (MIM-LUT) decoders, since they adopt a series of cascaded two-input single-output LUTs for decoding.
However, concatenating LUTs causes the degradation of both mutual information and error rate performance.

To overcome the drawbacks of the FAIDs in \cite{Richardson01capacity, Lee05, Thorpe02, Romero16, Lewandowsky18, meidlinger2020design}, we have presented a mutual information-maximizing quantized belief propagation (MIM-QBP) decoder in \cite{he2019onmutual}.
The MIM-QBP decoder adopts the RCQ decoding architecture at both the CN and VN update similarly to the FAIDs in  \cite{Richardson01capacity, Lee05, Thorpe02}.
The main contribution of \cite{he2019onmutual} is to establish a high-level principle for designing good RCQ parameters, aiming at maximizing  the MI between the coded bits and the exchanged messages, for decoding any regular LDPC codes.
The principle has enabled the development of many follow-up MIM-FAIDs \cite{Wang2022rcq,Mohr2021iblayer,Kang2022qms, kang2022generalized,Kang2022qsms,lv2022qlms}.
These MIM-FAIDs' VN update follows the RCQ decoding architecture where the RCQ parameters are designed based on the principle established in \cite{he2019onmutual};
Their CN update employs the $\min$ operation (like the min-sum decoder \cite{Chen05}) to simplify the RCQ decoding architecture used in the CN update of the MIM-QBP decoder, at the cost of slightly degrading error rate performance.

This paper, as the extended version of \cite{he2019onmutual}, presents the first complete and systematic design framework for the RCQ parameters of the MIM-QBP decoder.
More specifically, Section \ref{section: MIM-QBP decoding} of this paper includes the design principle originally proposed in \cite{he2019onmutual} and gives more explanations.
The most important extension over \cite{he2019onmutual} is that, this paper contains a new section (Section \ref{section: design of MIM-QBP decoder}) to illustrate the optimality of a class of  reconstruction functions (RFs) which can make the corresponding MIM-QBP decoder maximize the MI between the coded bits and the exchanged messages;
Meanwhile, it presents a near-optimal design of practical RFs based on scaling the optimal RFs.
Therefore, this paper essentially establishes a fundamental theory for supporting the validity of the aforementioned MIM-FAIDs \cite{Wang2022rcq,Mohr2021iblayer,Kang2022qms, kang2022generalized,Kang2022qsms,lv2022qlms}.
Moreover, we believe that our design framework will facilitate the development of more follow-up MIM-FAIDs.
The main contributions of this paper are summarized as follows:
\begin{itemize}
\item   This paper presents, for the first time, a complete and systematic design framework of the RCQ parameters for decoding any regular LDPC codes.
\item   We prove that with sufficient precision for node update, our design for the RCQ parameters is able to maximize the MI between the coded bits and the exchanged messages. 
\item   We investigate the error rate performance of the proposed MIM-QBP decoder based on extensive simulations. We observe that the MIM-QBP decoders can always considerably outperform the MIM-LUT decoders, and can sometimes even outperform the floating-point BP decoder with only 3 bits for each exchanged message.
\end{itemize}

The remainder of this paper is organized as follows.
Section \ref{section: preliminaries} first introduces the optimal quantization method for the binary-input discrete memoryless channel (DMC), and then gives a review of the MIM-LUT decoding and also highlights the linkage between the two topics.
Section \ref{section: MIM-QBP decoding} illustrates the general design principle for designing the RCQ parameters for decoding  regular LDPC codes.
Section \ref{section: design of MIM-QBP decoder} develops an efficient design of the RFs of the MIM-QBP decoder.
Section \ref{section: simulation results} presents the simulation results.
Finally, Section \ref{section: conclusion} concludes this paper.

\section{Preliminaries}
\label{section: preliminaries}

\subsection{MIM Quantization of Binary-Input DMC}\label{section: DP quantization}

\begin{figure}[t]
\centering
\includegraphics[scale = 0.5]{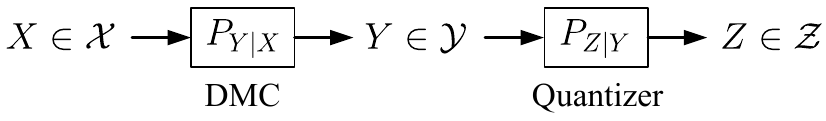}
\caption{Quantization of a discrete memoryless channel (DMC).}
\label{fig: DMC}
\end{figure}

Due to the strong linkage between the MIM based channel quantization and the MIM based LDPC decoding message quantization, we first review the quantization of a binary-input DMC.
As shown by Fig. \ref{fig: DMC}, the channel input $X$ takes values from $\mc{X} = \{0, 1\}$ with probability $P_X(0)$ and $P_X(1)$, respectively.
The channel output $Y$ takes values from $\mc{Y} = \{y_1, y_2, \ldots, y_N\}$ with channel transition probability given by $P_{Y|X}(y_j | x) = Pr(Y = y_j | X = x)$, where $x = 0, 1$ and $j = 1, 2, \ldots, N$.
The channel output $Y$ is quantized to $Z$ which takes values from $\mc{Z} = \{1, 2, \ldots, M\}$.
A well-known criterion for channel quantization \cite{Kurkoski14, he2021dynamic} is to design a quantizer $Q^*: \mc{Y} \to {Z}$ to maximize the MI between $X$ and $Z$, i.e.
\begin{align}\label{eqn: MI quantizer Q*}
    Q^* &= \arg \max_{Q} I(X; Z)\nonumber\\
    &= \arg \max_{Q} \sum_{x \in \mc{X}, z \in \mc{Z}} P_{X,Z}(x,z) \log \frac{P_{X,Z}(x,z)}{P_{X}(x) P_{Z}(z)},
\end{align}
where $P_{X, Z}(x, z) = P_{X}(x) \sum_{y \in \mc{Y}} P_{Y|X}(y|x) P_{Z|Y}(z|y)$ and $P_Z(z) = \sum_{x \in \mc{X}} P_{X, Z}(x, z)$.

A deterministic quantizer (DQ) $Q: \mc{Y} \to \mc{Z}$ means that for each $y \in \mc{Y}$, there exists a unique $z \in \mc{Z}$ such that $P_{Z|Y}(z|y) = 1$ and $P_{Z|Y}(z'|y) = 0$ for $z \neq z' \in \mc{Z}$.
Let $Q^{-1}(z) \subset \mc{Y}$ denote the preimage of $z \in \mc{Z}$.
We name $Q$ a sequential deterministic quantizer (SDQ) \cite{he2021dynamic} if it can be equivalently described by an integer set $\Lambda = \{\lambda_0, \lambda_1, \ldots, \lambda_{M-1}, \lambda_M\}$ with $\lambda_0 = 0 < \lambda_1 < \cdots < \lambda_{M-1} < \lambda_M = N$ in the way given below
\begin{equation*}
\left\{
\begin{array}{l}
    Q^{-1}(1) \,\,\,= \{{y_1, y_2, \ldots, y_{\lambda_1}}\},\\
    Q^{-1}(2) \,\,\,= \{y_{\lambda_1 + 1}, y_{\lambda_1 + 2},  \ldots, y_{\lambda_2}\},\\
    \quad\quad\quad\quad\vdots\\
    Q^{-1}(M) = \{y_{\lambda_{M-1} + 1}, y_{\lambda_{M-1} + 2},  \ldots, y_{\lambda_M}\}.
\end{array}
\right.
\end{equation*}
We thus also name $\Lambda$ an SDQ.

According to \cite{Kurkoski14}, there always at least exists a deterministic $Q^*$ for  \eqref{eqn: MI quantizer Q*} that can maximize $I(X; Z)$.
Moreover, $Q^*$ is an optimal SDQ if $P_{Y|X}$ further satisfies
\begin{equation}\label{eqn: LLR increasing}
\frac{P_{Y|X}(y_1|0)}{P_{Y|X}(y_1|1)} \geq \frac{P_{Y|X}(y_2|0)}{P_{Y|X}(y_2|1)} \geq \cdots \geq \frac{P_{Y|X}(y_{N}|0)}{P_{Y|X}(y_N|1)},
\end{equation}
where if $P_{Y|X}(\cdot|1) = 0$, we regard $\frac{P_{Y|X}(\cdot|0)}{P_{Y|X}(\cdot|1)} = \infty$ as the largest value. (However, $Q^*$ may not be an SDQ when \eqref{eqn: LLR increasing} does not hold, in which case $Q^*$ may not be simply described by the set $\Lambda$.)
Note that after merging any two elements $y, y' \in \mc{Y}$ with $P_{Y|X}(y|0)/{P_{Y|X}(y|1)} = P_{Y|X}(y'|0)/{P_{Y|X}(y'|1)}$, the resulting optimal quantizer is as optimal as the original one \cite{Kurkoski14,he2021dynamic}.
A general framework has been developed in \cite{he2021dynamic} for applying dynamic programming (DP) \cite[Section 15.3]{IntroAlgo01} to find an optimal SDQ among all SDQs to maximize $I(X; Z)$. (The condition of \eqref{eqn: LLR increasing} is not required, and if it holds, the optimal SDQ is an optimal DQ which can maximize $I(X; Z)$ among all quantizers from $\mc{Y}$ to $\mc{Z}$.)

\subsection{MIM-LUT Decoder Design for Regular LDPC Codes}
\label{section: MIM-LUT}

Consider a binary-input DMC.
Denote the channel input by $X$ which takes values from $\mc{X} = \{0, 1\}$ with equal probability, i.e., $P_X(0) = P_X(1) = 1/2$.
Denote $L$ as the DMC output which takes values from $\mc{L} = \{0, 1, \ldots, |\mc{L}|-1\}$ with channel transition probability $P_{L|X}$.
%
Consider the design of a quantized message passing (MP) decoder for a regular $(d_v, d_c)$ LDPC code, where $d_v$ and $d_c$ are the degrees of VNs and CNs, respectively.
Denote $\mc{R} = \{0, 1, \ldots, |\mc{R}|-1\}$ and $\mc{S} = \{0, 1, \ldots, |\mc{S}|-1\}$ as the alphabets of variable-to-check  (V2C) and check-to-variable (C2V) messages, respectively.
Note that $\mc{L}, \mc{R}, \mc{S}$ and their related functions may or may not vary with iterations.
We use these notations without specifying the associated iterations since after specifying the decoder design for one iteration, the design is clear for all the other iterations.

For the V2C message $R \in \mc{R}$ (resp. C2V message $S \in \mc{S}$), we use $P_{R|X}$ (resp. $P_{S|X}$) to denote the probability mass function (pmf) of $R$ (resp. $S$) conditioned on the channel input bit $X$.
If the code graph is cycle-free, $R$ (resp. $S$) conditioned on $X$ is independent and identically distributed (i.i.d.) with respect to different edges for a given iteration.
In the following, we introduce the design of the MIM-LUT decoder \cite{ Romero16,  meidlinger2020design, Lewandowsky18} based on density evolution (DE) \cite{Richardson01capacity}.
In particular, DE is carried out to construct the LUTs of the message mappings for decoding by tracking the pmfs $P_{R|X}$ and $P_{S|X}$ at each iteration.
Note that although a cycle-free code graph is assumed in the design of the MIM-LUT decoder, it still works well on code graphs containing cycles as shown in \cite{ Romero16,  meidlinger2020design,  Lewandowsky18}.

\begin{figure}[t]
\centering
\includegraphics[scale = 0.5]{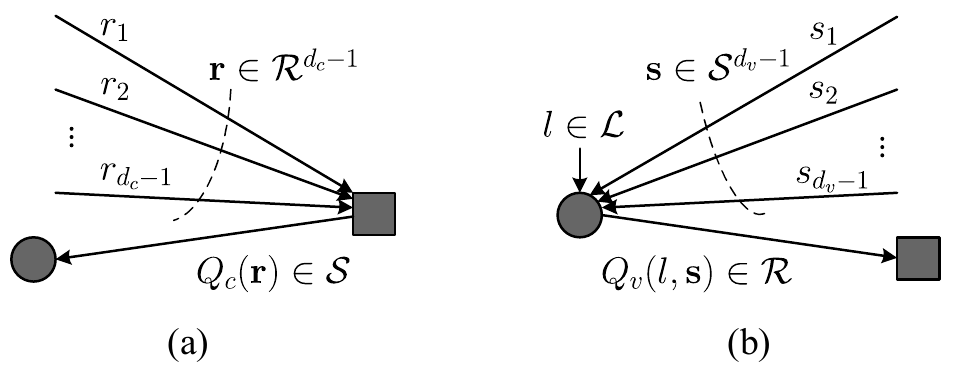}
\caption{Node update for mutual information-maximizing lookup table (MIM-LUT) decoding, where the circle and square represent a variable and check node, respectively. (a) Check node update. (b) Variable node update.}
\label{fig: node update}
\end{figure}

\begin{figure}[t]
\centering
\includegraphics[scale = 0.5]{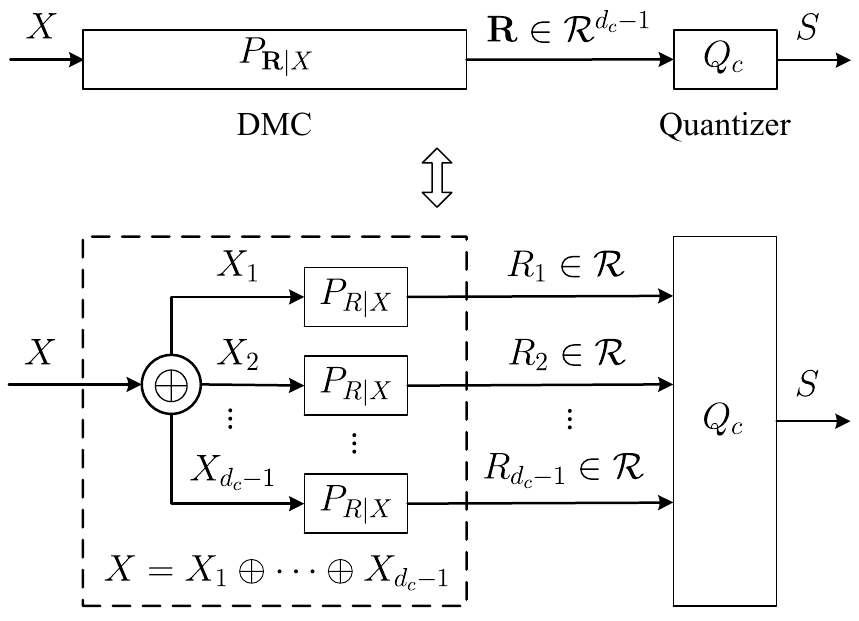}
\caption{Quantization of a discrete memoryless channel (DMC), where the quantizer works exactly the same as the check node update function $Q_c$ for the mutual information-maximizing lookup table (MIM-LUT) decoding shown by Fig. \ref{fig: node update}(a).}

\label{fig: CN_update_channel}
\end{figure}

For each iteration, we first design the update function (UF)
\begin{equation}\label{eqn: def of Q_c}
    Q_{c}: \mc{R}^{d_c - 1} \to \mc{S}
\end{equation}
for the CN update, which is shown by Fig. \ref{fig: node update}(a).
The MIM-LUT decoding methods design $Q_{c}$ to maximize $I(X; S)$.
For easy understanding, we can equivalently convert it to the problem of  DMC quantization, as shown by Fig. \ref{fig: CN_update_channel}.

We assume $P_{R|X}$ is known, since for the first iteration, $P_{R|X}$ can be solely derived from the channel transition probability $P_{L|X}$, and for the other iteration, $P_{R|X}$ is known after the design at VN is completed.
The joint distribution $P_{\mb{R}|X}$ of the incoming message $\mb{R} \in \mc{R}^{d_c - 1}$ conditioned on the channel input bit $X$ (i.e., the channel transition probability  $P_{\mb{R}|X}$ of the DMC shown by Fig. \ref{fig: CN_update_channel}) is given by \cite{Romero16}
\begin{equation}\label{eqn: joint P_R|X}
    P_{\mb{R}|X}(\mb{r}|x) = \left(\frac{1}{2}\right)^{\dim(\mb{r}) - 1} \sum_{\mb{x}: \oplus \mb{x} = x} \prod_{i = 1}^{\dim(\mb{r})} P_{R|X}(r_i|x_i),
\end{equation}
where $\mb{r} = (r_1, r_2, \ldots, r_{d_c - 1}) \in \mc{R}^{d_c - 1}$ is a realization of $\mb{R}$, $\dim(\mb{r}) = d_c - 1$ is the dimension of $\mb{r}$, $x \in \mc{X}$ is a realization of $X$, $\mb{x} = (x_1, x_2, \ldots, x_{d_c - 1}) \in \mc{X}^{d_c - 1}$ consists of channel input bits corresponding to the VNs associated with incoming edges, and $\oplus \mb{x} = x_1 \oplus x_2 \oplus \cdots \oplus x_{d_c - 1}$ with $\oplus$ denoting the addition in $GF(2)$.
Based on \eqref{eqn: joint P_R|X}, we have
\begin{equation}\label{eqn: P(R|0) - P(R|1)}
    \left\{
    \begin{array}{l}
        P_{\mb{R}|X}(\mb{r}|0) \pm P_{\mb{R}|X}(\mb{r}|1) = \left(\frac{1}{2}\right)^{\dim(\mb{r}) - 1} \times \\
        \quad\quad\quad\quad\quad   \prod_{i=1}^{\dim(\mb{r})} (P_{R|X}(r_i | 0) \pm P_{R|X}(r_i | 1)),\\
        P_{X|\mb{R}}(0|\mb{r}) \pm P_{X|\mb{R}}(1|\mb{r}) = \\
        \quad\quad\quad\quad\quad \prod_{i=1}^{\dim(\mb{r})} (P_{X|R}(0 | r_i) \pm P_{X|R}(1 | r_i)).
    \end{array}
    \right.
\end{equation}

Given $P_{\mb{R}|X}$, the design of $Q_c$ is equivalent to the design of $Q^*$ in \eqref{eqn: MI quantizer Q*} by setting $\mc{Y} = \mc{R}^{d_c - 1}$ and $\mc{Z} = \mc{S}$.
We can solve this design problem by using the DP method proposed in \cite{Kurkoski14}, after listing $\mb{r}$ in descending order based on $P_{\mb{R}|X}(\mb{r}|0) / P_{\mb{R}|X}(\mb{r}|1)$ (see \eqref{eqn: LLR increasing}).
After designing $Q_{c}$, the output message $S$ is passed to the CN's neighbour VNs, with $P_{S|X}$ being given by
\begin{equation}\label{eqn: P(S|X)}
    P_{S|X}(s|x) = \sum_{\mb{r} \in Q_c^{-1}(s)} P_{\mb{R} | X}(\mb{r} | x).
\end{equation}

\begin{figure}[t]
\centering
\includegraphics[scale = 0.48]{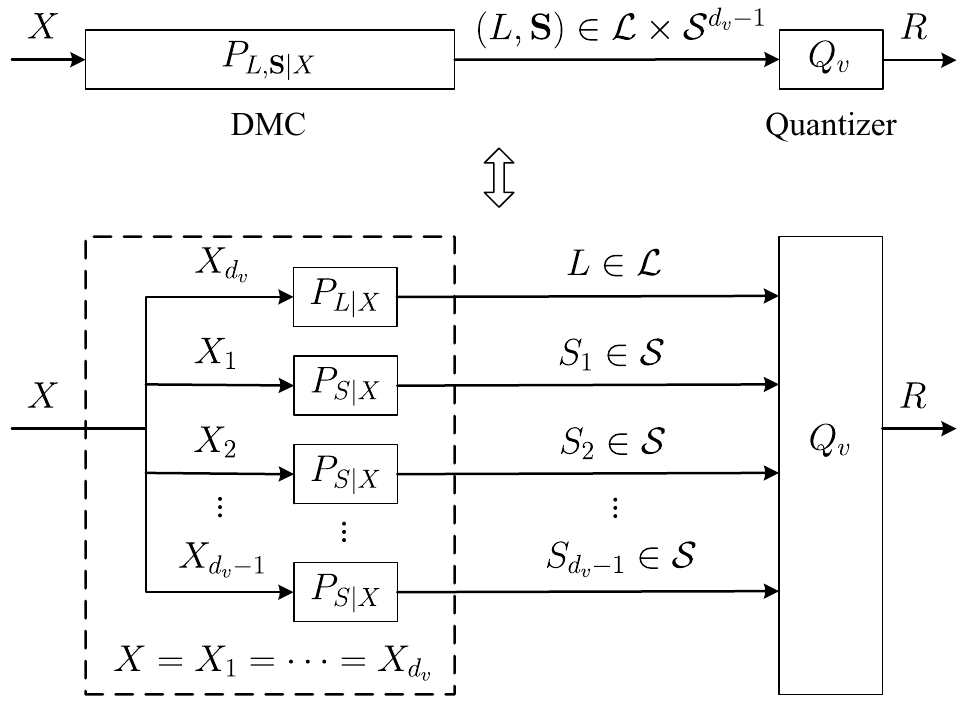}
\caption{Quantization of a discrete memoryless channel (DMC), where the quantizer works exactly the same as the variable node update function $Q_v$ for the mutual information-maximizing lookup table (MIM-LUT) decoding shown by Fig. \ref{fig: node update}(b).}

\label{fig: VN_update_channel}
\end{figure}

We then proceed to design the UF
\begin{equation}\label{eqn: def of Q_v}
    Q_{v}: \mc{L} \times \mc{S}^{d_v - 1} \to \mc{R}
\end{equation}
for the VN update, which is shown by Fig. \ref{fig: node update}(b).
The MIM-LUT decoding methods also design $Q_{v}$ to maximize $I(X; R)$.
For easy understanding, we can equivalently convert it to the problem of  DMC quantization, as shown by Fig. \ref{fig: VN_update_channel}.

The joint distribution $P_{L,\mb{S}|X}$ of incoming message $(L, \mb{S}) \in \mc{L} \times \mc{S}^{d_v - 1}$ conditioned on the channel input bit $X$ (i.e., the channel transition probability  $P_{L, \mb{S}|X}$ of the DMC shown by Fig. \ref{fig: VN_update_channel}) is given by\cite{Romero16}
\begin{equation}\label{eqn: joint P_L,S|X}
    P_{L,\mb{S}|X}(l, \mb{s}|x) =  P_{L|X}(l|x) \prod_{i = 1}^{\dim(\mb{s})} P_{S|X}(s_i|x),
\end{equation}
where $l \in \mc{L}$ is a realization of $L$, $\mb{s} = (s_1, s_2, \ldots, s_{d_v - 1}) \in \mc{S}^{d_v - 1}$ is a realization of $\mb{S}$, $\dim(\mb{s}) = d_v - 1$ is the dimension of $\mb{s}$, and $x \in \mc{X}$ is a realization of $X$.

Given $P_{L, \mb{S}|X}$, the design of $Q_{v}$ is equivalent to the design of $Q^*$ in \eqref{eqn: MI quantizer Q*} by setting $\mc{Y} = \mc{L} \times \mc{S}^{d_v - 1}$ and $\mc{Z} = \mc{R}$.
We can solve this design problem by using the DP method proposed in \cite{Kurkoski14}, after listing $(l, \mb{s})$ in descending order based on $P_{L,\mb{S}|X}(l, \mb{s}|0) / P_{L,\mb{S}|X}(l, \mb{s}|1)$ (see \eqref{eqn: LLR increasing}).
After designing $Q_{v}$, the output message $R$ is passed to the VN's neighbour CNs, with $P_{R|X}$ given by
\begin{equation}\label{eqn: P(R|X)}
    P_{R|X}(r|x) = \sum_{(l, \mb{s}) \in Q_v^{-1}(r)} P_{L, \mb{S} | X}(l, \mb{s} | x).
\end{equation}

For each iteration, we can design the estimation function
\begin{equation}\label{eqn: def of Q_e}
    Q_{e}: \mc{L} \times \mc{S}^{d_v} \to \mc{X}
\end{equation}
to estimate the channel input bit corresponding to each VN.
The design of $Q_e$ can be carried out similarly to that of $Q_v$.
The main differences involved in the design lie in the aspect that i) the incoming message alphabet $\mc{L} \times \mc{S}^{d_v - 1}$ is changed to $\mc{L} \times \mc{S}^{d_v}$; and ii) the outgoing message alphabet $\mc{R}$ is changed to $\mc{X}$.
We thus ignore the details.

After completing the design of $Q_c$, $Q_v$, and $Q_e$ for all iterations, the design of the MIM-LUT decoder is completed.
In general, $|\mc{L}| = |\mc{R}| = |\mc{S}| = 8$ (resp. $16$) is used for all iterations, leading to a 3-bit (resp. 4-bit) decoder.
Given $|\mc{L}|, |\mc{R}|, |\mc{S}|$, and the maximum allowed decoding iterations $I_\text{max}$, the design of the MIM-LUT decoder is determined by $P_{L|X}$.
As shown in the literature \cite{ Romero16,  meidlinger2020design,  Lewandowsky18}, $P_{L|X}$ depends on the choice of the noise standard derivation $\sigma_d$ for designing the decoder parameters for an AWGN channel.
We define the design threshold of an MIM-LUT decoder as
\begin{equation}
\label{sigma}
{\sigma ^*} = \sup \left\{ {\sigma_d :{I^{({I_{\text{max} }})}}(X;R) > 1 -\xi} \right\},
\end{equation}
where $I^{({I_{\text{max} }})}(X;R)$ is the value of $I(X;R)$ computed after $I_\text{max}$ iterations, and $\xi$ is a preset small number (e.g. $10^{-2}$).
Based on extensive simulations, by selecting $\sigma_d=\sigma^*$, the MIM-LUT decoder can achieve very good error rate performance across a wide range of signal-to-noise ratios (SNRs).
However, the underlying reason remains an open problem.

Note that $Q_{c}$, $Q_{v}$, and $Q_{e}$ are stored as LUTs when implementing the MIM-LUT decoding.
The sizes of the tables for $Q_{c}$, $Q_{v}$, and $Q_{e}$ are $|\mc{R}|^{d_c - 1}$, $|\mc{L}| \cdot |\mc{S}|^{d_v - 1}$, and $|\mc{L}| \cdot |\mc{S}|^{d_v}$, respectively.
Thus, a huge memory requirement may arise when the sizes of the tables are large in practice.
To solve this problem, current MIM-LUT decoding methods \cite{ Romero16,  Lewandowsky18,  meidlinger2020design} need to decompose $Q_{c}$, $Q_{v}$, and $Q_{e}$ into a series of two-input single-output LUTs.
The decomposition can significantly reduce the cost of storage.
However, it will degrade the performance of $Q_{c}$, $Q_{v}$, and $Q_{e}$ compared to the case without decomposition.

\section{MIM-QBP Decoding for Regular LDPC Codes}
\label{section: MIM-QBP decoding}

In this section, we propose a general design framework for the MIM-QBP decoder.
This decoder follows the RCQ decoding architecture and can be implemented based only on simple mappings and additions.
Moreover, it can handle all incoming messages at a given node (CN or VN) simultaneously without causing any storage problem.
As a result, the MIM-QBP decoder can greatly reduce the memory consumption and avoid the error rate performance loss due to table decomposition compared to the MIM-LUT decoder.
In the following, we specify the design principles of the RCQ parameters at CN and VN, respectively.

\subsection{CN Update for MIM-QBP Decoding}\label{section: MIM-QBP decoding at CN}

\begin{figure}[t]
\centering
\includegraphics[scale = 0.5]{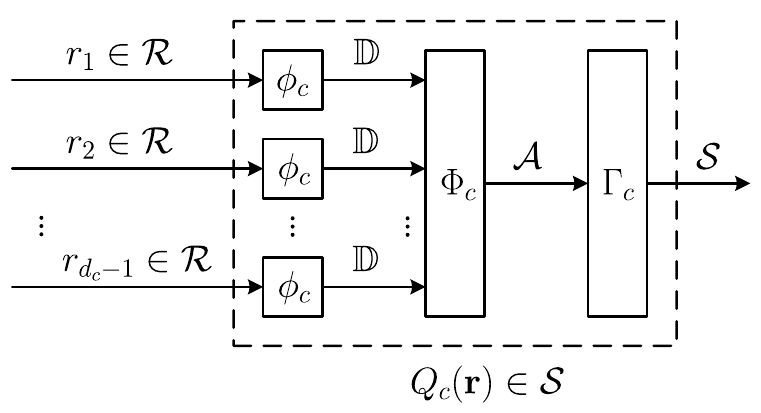}
\caption{Check node update for mutual information-maximizing quantized belief propagation (MIM-QBP) decoding. The part enclosed by the dash square corresponds to the update operation in the CN of Fig. \ref{fig: node update}(a).}

\label{fig: CN_update_MIMQBP}
\end{figure}

As shown by Fig. \ref{fig: CN_update_MIMQBP}, the CN update for the MIM-QBP decoding proceeds the following three steps (reconstruction, calculation, and quantization):
First, an RF $\phi_c$ is used to map each V2C message symbol to a  number of larger bit width which is called V2C computational message;
second, we use a function $\Phi_c$ to combine all V2C computational messages  together as defined by \eqref{eqn: def of Phi_c} to form the C2V computational message;
third, we use an SDQ $\Gamma_c$ to quantize the C2V computational message into the C2V message symbol.
In this way, the UF $Q_c$ at the CN is fully determined by $\phi_c, \Phi_c$, and $\Gamma_c$.
In the rest of this subsection, we show the principles of designing $\phi_c, \Phi_c$, and $\Gamma_c$ so as to obtain a $Q_c$ that can maximize $I(X; S)$.

\subsubsection{Design principle of $\phi_c$}
We use an RF
\begin{equation}\label{eqn: def of phi_c}
    \phi_c: \mc{R} \to \mathbb{D}
\end{equation}
to map each V2C message realization $r \in \mc{R}$ to a computational message $\phi_c(r)$ in the computational domain $\mathbb{D}$, where we set $\mathbb{D} = \mathbb{R}$ (real number set) or $\mathbb{D} = \mathbb{Z}$ (integer set) for different considerations.
Let $\mr{sgn}(\alpha)$ be the sign of $\alpha \in \mathbb{R}$, and
\[
    \mr{sgn}(\alpha) =
    \begin{cases}
        -1 & \alpha < 0,\\
        0 & \alpha = 0,\\
        1 & \alpha > 0.
    \end{cases}
\]
For $r \in \mc{R}$, let
\[
    LLR(r) = \log \left( P_{X|R}(0|r) / P_{X|R}(1|r) \right).
\]
A good choice for $\phi_c(r)$ based on empirical observation is
\begin{align}\label{eqn: require phi_c}
\left\{
    \begin{array}{l}
    \mr{sgn}(\phi_c(r)) = \mr{sgn}(LLR(r)),\\
    |\phi_c(r)| \propto \frac{1}{|LLR(r)|}.
    \end{array}
\right.
\end{align}
Based on (\ref{eqn: require phi_c}), we associate $\phi_c(r)$ to the channel input bit $X$ in the following way:
we predict $X$ to be 0 if $\mr{sgn}(\phi_c(r)) > 0$ and to be 1 if $\mr{sgn}(\phi_c(r)) < 0$, while $|\phi_c(r)|$ indicates the \emph{unreliability} of the prediction result (larger $|\phi_c(r)|$ means less reliability).
Note that the intuition behind \eqref{eqn: require phi_c} comes from \cite{Gallager62} and \cite{Chen05}.
In \cite{Gallager62} and \cite{Chen05}, the function $f(\theta) = \log((e^\theta + 1) / (e^\theta - 1))$ is adopted by the CN update for any positive LLR value $\theta$, where $f(\theta)$ decreases monotonically with $\theta$.
\subsubsection{Design principle of $\Phi_c$}
For each V2C message realization $\mb{r} \in \mc{R}^{d_c - 1}$, we combine all the V2C computational messages to form the C2V computational message as follows:
\begin{equation}\label{eqn: def of Phi_c}
    \Phi_c(\mb{r}) = \left( \prod_{i = 1}^{\dim(\mb{r})} \mr{sgn}(\phi_c(r_i)) \right) \sum_{i = 1}^{\dim(\mb{r})} |\phi_c(r_i)|.
\end{equation}
Similar to the CN update employed in \cite{Gallager62} and \cite{Chen05}, we predict $X$ to be 0 if $\mr{sgn}(\Phi_c(\mb{r})) = \prod_{i = 1}^{\dim(\mb{r})} \mr{sgn}(\phi_c(r_i)) > 0$, and to be 1 if $\mr{sgn}(\Phi_c(\mb{r})) < 0$.
Note that $|\Phi_c(\mb{r})| = \sum_{i = 1}^{\dim(\mb{r})} |\phi_c(r_i)|$ indicates the \emph{unreliability} of the prediction result.
Prediction in this way is consistent with the true situation shown by Fig. \ref{fig: CN_update_channel}: $X$ is the binary summation of the channel input bits associated with $\mb{r}$, which is determined by $\mr{sgn}(\phi_c(r_i)), i = 1, 2, \ldots, d_c - 1$.
More incoming messages lead to more unreliability (i.e., larger $\dim(\mb{r})$ leads to larger $|\Phi_c(\mb{r})|$. This is the reason why we use $|\phi_c(r)|$ as the unreliability.).
The calculation step by \eqref{eqn: def of Phi_c} results in a set of distinct values of $\Phi_c(\mb{r}), \forall\, \mb{r} \in \mc{R}^{d_c - 1}$, which we denote by
\begin{equation}\label{eqn: def of mc_A}
    \mc{A} = \{a_1, a_2, \ldots, a_{|\mc{A}|}\}.
\end{equation}
The elements in $\mc{A}$ are labelled to satisfy
\begin{equation}\label{eqn: order of A}
    a_1 \succ a_2 \succ \cdots \succ  a_{|\mc{A}|},
\end{equation}
where $\succ$ is a binary relation on $\mathbb{R}$ defined by
\begin{align*}
    \alpha \succ \beta \iff &\mr{sgn}(\alpha) > \mr{sgn}(\beta) \text{~or~} \\
    &(\mr{sgn}(\alpha) = \mr{sgn}(\beta) \text{~and~} \alpha < \beta)
\end{align*}
for $\alpha, \beta \in \mathbb{R}$.
For example, we have $1 \succ 2 \succ 0 \succ -2 \succ -1$.
Assuming $\Phi_c(\mb{r}) = a_i$, we know from \eqref{eqn: order of A} that it is more likely to predict $X$ to be 0 for smaller $i$ and to be 1 for larger $i$.
Thus, the listing order of \eqref{eqn: order of A} has a similar feature as that of \eqref{eqn: LLR increasing}.
Let $A$ be a random variable taking values from $\mc{A}$.
With \eqref{eqn: def of Phi_c}, the pmf of $A$ conditioned on the channel input bit $X$ is
\begin{equation}\label{eqn: prob A|X}
    P_{A|X}(a_i|x) = \sum_{\mb{r} \in \mc{R}^{d_c - 1}, \Phi_c(\mb{r}) = a_i} P_{\mb{R} | X}(\mb{r} | x),
\end{equation}
where $1 \leq i \leq |\mc{A}|$, and $P_{\mb{R} | X}(\mb{r} | x)$ is given by \eqref{eqn: joint P_R|X}.

\subsubsection{Design principle of $\Gamma_c$}
Based on $\mc{A}$ and $P_{A|X}$, we adopt the general DP method proposed in \cite{he2021dynamic} to find an optimal SDQ
\begin{equation}\label{eqn: def of Lambda_c}
    \Lambda_c = \{\lambda_0 = 0, \lambda_1, \ldots, \lambda_{|\mc{S}|-1}, \lambda_{|\mc{S}|} = |\mc{A}|\}: \mc{A} \to \mc{S}
\end{equation}
to maximize $I(X; S)$ among all SDQs.
Here $\lambda_1, \lambda_2, \ldots, \lambda_{|\mc{S}|-1}$ are corresponding to the indices of $|\mc{S}|-1$ elements in $\mc{A}$.
Based on $\Lambda_c$, instead of using \eqref{eqn: P(S|X)}, we can compute $P_{S|X}$ for the C2V message $S$ in a simpler way given by
\begin{equation}\label{eqn: P_(S|X) lambda}
P_{S|X}(s|x) = \sum_{i = \lambda_s + 1}^{\lambda_{s+1}} P_{A|X}(a_i|x).
\end{equation}
Since the indices in $\Lambda_c$ cannot be directly applied to decoding, we convert the index $\lambda_i \in \Lambda_c$ ($1 \le i \le |\mc{S}|-1$) to the $\lambda_i$-th element $a_{{\lambda _i}} \in \mc{A}$, which leads to the threshold set (TS) $\Gamma_c$ given by
\begin{equation}\label{eqn: def of Gamma_c}
    \Gamma_c = \{\gamma_i=a_{\lambda_i}: 1 \le i \le |\mc{S}|-1 \}.
\end{equation}

At this point, the UF $Q_c: \mc{R}^{d_c - 1} \to \mc{S}$ is fully determined by $\phi_c, \Phi_c$, and $\Gamma_c$ in the following way as
\begin{equation}\label{eqn: def of Q_c by Gamma}
    Q_c(\mb{r}) =
    \begin{cases}
        0 & \Phi_c(\mb{r}) \succeq \gamma_1,\\
        i &  \gamma_{i} \succ \Phi_c(\mb{r}) \succeq \gamma_{i+1}, 1 \leq i \leq |\mc{S}| - 2,\\
        |\mc{S}| - 1 & \gamma_{|\mc{S}| - 1} \succ \Phi_c(\mb{r}),\\
    \end{cases}
\end{equation}
where $\succeq$ is a binary relation on $\mathbb{R}$ defined by
\[
    \alpha \succeq \beta \iff \alpha \succ \beta \text{~or~} \alpha = \beta, \text{~for~} \alpha, \beta \in \mathbb{R}.
\]

Note that the storage complexity of $Q_c$ given by \eqref{eqn: def of Q_c by Gamma} is $O(|\mc{R}| + |\mc{S}|)$ ($O(|\mc{R}|)$ for storing $\phi_c$ and $O(|\mc{S}|)$ for storing $\Gamma_c$), which is negligible since each element of $\phi_c$ and $\Gamma_c$ is a small integer in practice.
On the other hand, implementing the CN update shown by Fig. \ref{fig: CN_update_MIMQBP} for \emph{one} outgoing message has complexity $O(d_c + \lceil \log_2(|\mc{S}|) \rceil)$.
In particular, computing $\Phi_c(\mb{r})$ has complexity $O(d_c)$ (binary operations mainly including additions), which allows a binary tree-like parallel implementation; meanwhile, mapping $\Phi_c(\mb{r})$ to $S$ based on $\Gamma_c$ has complexity $O(\lceil \log_2(|\mc{S}|) \rceil)$ (binary comparison operations).
Note that the simple implementation for mapping $\Phi_c(\mb{r})$ to $S$ indeed benefits from the use of SDQ rather than the optimal DQ.
Because if an optimal DQ is used to map $\Phi_c(\mb{r}) \in \mc{A}$ to $\mc{S}$ in \eqref{eqn: def of Lambda_c}, it requires an additional table of size $|\mc{A}|$ to store this optimal DQ.
This is the essential reason why we choose SDQs in the quantization step.

\subsection{VN Update for MIM-QBP Decoding}\label{section: MIM-QBP decoding at VN}

\begin{figure}[t]
\centering
\includegraphics[scale = 0.5]{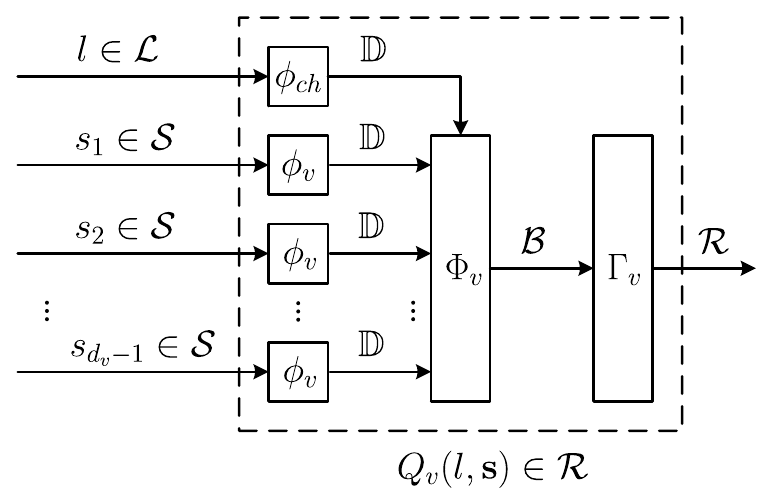}
\caption{Variable node update for mutual information-maximizing quantized belief propagation (MIM-QBP) decoding. The part enclosed by the dash square corresponds to the update operation in the VN of Fig. \ref{fig: node update}(b).}

\label{fig: VN_update_MIMQBP}
\end{figure}

As shown by Fig. \ref{fig: VN_update_MIMQBP}, the VN update for MIM-QBP decoding consists of the following three steps (reconstruction, calculation, and quantization): First, we use RF $\phi_v$ (resp. $\phi_{ch}$) to map each C2V (resp. channel) message symbol  to a number of larger bit width which is called C2V (channel-to-variable) computational message;
second, we use a function $\Phi_v$ to  combine all computational messages  together as defined by \eqref{eqn: def of Phi_v} to form the V2C computational message;
third, we use an SDQ $\Gamma_v$ to quantize the V2C computational message into the V2C message symbol.
In this way, the UF $Q_v$ at the VN is fully determined by $\phi_v, \phi_{ch}, \Phi_v,$ and $\Gamma_v$.
In the rest of this subsection, we show the principles of designing $\phi_v, \phi_{ch}, \Phi_v,$ and $\Gamma_v$ so as to obtain a $Q_v$ that can maximize $I(X; R)$.

\subsubsection{Design principle of $\phi_v$ and $\phi_{ch}$}
We use an RF
\begin{equation}\label{eqn: def of phi_v}
    \phi_v: \mc{S} \to \mathbb{D}
\end{equation}
to map each C2V message realization $s \in \mc{S}$ to $\phi_v(s) \in \mathbb{D}$, and use another RF
\begin{equation}\label{eqn: def of phi_ch}
    \phi_{ch}: \mc{L} \to \mathbb{D}
\end{equation}
to map the channel output message realization $l \in \mc{L}$ to $\phi_{ch}(l) \in \mathbb{D}$.
For $s \in \mc{S}$, let
\[
    LLR(s) = \log\left( P_{X|S}(0|s) / P_{X|S}(1|s) \right).
\]
For $l \in \mc{L}$, let
\[
    LLR(l) = \log\left( P_{X|L}(0|l) / P_{X|L}(1|l) \right).
\]
A good choice for $\phi_v(s)$ and $\phi_{ch}(l)$ based on empirical observation is
\begin{align}\label{eqn: require phi_v}
\left\{
    \begin{array}{l}
    \phi_v(s) \propto LLR(s),\\
    \phi_{ch}(l) \propto LLR(l).
    \end{array}
\right.
\end{align}
Based on (\ref{eqn: require phi_v}), we associate $\phi_v(s)$ and $\phi_{ch}(l)$ to the channel input bit $X$ in the following way:
$X$ is more likely to be 0 (resp. 1) for larger (resp. smaller) $\phi_v(s)$ and $\phi_{ch}(l)$.

\subsubsection{Design principle of $\Phi_v$}
For each incoming message realization  $(l, \mb{s}) \in \mc{L} \times \mc{S}^{d_v - 1}$, we combine the corresponding incoming computational messages to form the V2C computational message by
\begin{equation}\label{eqn: def of Phi_v}
    \Phi_v(l, \mb{s}) = \phi_{ch}(l) + \sum_{i = 1}^{\dim(\mb{s})} \phi_v(s_i).
\end{equation}
The channel input bit $X$ is more likely to be 0 (resp. 1) for smaller (resp. larger) $\Phi_v(l, \mb{s})$.
The calculation step by (\ref{eqn: def of Phi_v}) results in a set of distinct values of $\Phi_v(l, \mb{s}), \forall \, (l, \mb{s}) \in \mc{L} \times \mc{S}^{d_v - 1}$, which we denote by
\begin{equation}\label{eqn: def of mc_B}
    \mc{B} = \{b_1, b_2, \ldots, b_{|\mc{B}|}\}.
\end{equation}
The elements in $\mc{B}$ are labelled to satisfy
\begin{equation}\label{eqn: order of B}
    b_{1} > b_2 > \cdots > b_{|\mc{B}|}.
\end{equation}
Assuming $\Phi_v(l, \mb{s}) = b_i$, we know from \eqref{eqn: order of B} that $X$ is more likely be 0 (resp. 1) for smaller (resp. larger) $i$.
Thus, the listing order of \eqref{eqn: order of B} has a similar feature as the listing order of \eqref{eqn: LLR increasing}.
Let $B$ be a random variable taking values from $\mc{B}$.
With \eqref{eqn: def of Phi_v}, the pmf of $B$ conditioned on the channel input bit $X$ is
\begin{equation}\label{eqn: prob B|X}
    P_{B|X}(b_i|x) = \sum_{(l, \mb{s}) \in \mc{L} \times \mc{S}^{d_v - 1}, \Phi_v(l, \mb{s}) = b_i} P_{L, \mb{S} | X}(l, \mb{s} | x),
\end{equation}
where $1 \leq i \leq |\mc{B}|$ and $P_{L, \mb{S} | X}(l, \mb{s} | x)$ is given by \eqref{eqn: joint P_L,S|X}.

\subsubsection{Design principle of $\Gamma_v$}
Based on $\mc{B}$ and $P_{B|X}$, we apply the general DP method in \cite{he2021dynamic} to find an optimal SDQ
\begin{equation}\label{eqn: def of Lambda_v}
    \Lambda_v = \{\lambda_0 = 0, \lambda_1, \ldots, \lambda_{|\mc{R}|-1}, \lambda_{|\mc{R}|} = |\mc{B}|\}: \mc{B} \to \mc{R}
\end{equation}
to maximize $I(X; R)$ among all SDQs.
Here $\lambda_1, \lambda_2, \ldots, \lambda_{|\mc{R}|-1}$ are corresponding to the indices of $|\mc{R}|-1$ elements in $\mc{B}$.
Based on $\Lambda_v$, instead of using \eqref{eqn: P(R|X)}, we can compute $P_{R|X}$ for the outgoing message $R$ in a simpler way given by
\begin{equation}\label{eqn: P_(R|X) lambda}
P_{R|X}(r|x) = \sum_{i = \lambda_r + 1}^{\lambda_{r+1}} P_{B|X}(b_i|x).
\end{equation}
Since the indices in $\Lambda_v$ cannot be directly applied to decoding, we convert the index $\lambda_i \in \Lambda_v$ ($1 \le i \le |\mc{R}|-1$) to the $\lambda_i$-th element $b_{{\lambda _i}} \in \mc{B}$, which leads to the threshold set (TS) $\Gamma_v$ given by
\begin{equation}\label{eqn: def of Gamma_v}
\Gamma_v = \{\gamma_i=b_{\lambda _i}: 1 \le i \le |\mc{R}|-1\}.
\end{equation}

At this point, the UF $Q_v: \mc{L} \times \mc{S}^{d_v - 1} \to \mc{R}$ is fully determined by $\phi_v, \phi_{ch}, \Phi_v$, and $\Gamma_v$ in the following way as
\begin{equation}\label{eqn: def of Q_v by Gamma}
    Q_v(l, \mb{s}) =
    \begin{cases}
        0 & \!\! \Phi_v(l, \mb{s}) \geq \gamma_1,\\
        i & \!\! \gamma_i > \Phi_v(l, \mb{s}) \geq \gamma_{i+1}, 1 \leq i \leq |\mc{R}| - 2,\\
        |\mc{R}| - 1 & \!\! \Phi_v(l, \mb{s}) < \gamma_{|\mc{R}| - 1}.\\
    \end{cases}
\end{equation}

Note that the storage complexity of $Q_v$ given by \eqref{eqn: def of Q_v by Gamma} is $O(|\mc{S}| + |\mc{L}| + |\mc{R}|)$ ($O(|\mc{S}|)$ for storing $\phi_v$, $O(|\mc{L}|)$ for storing $\phi_{ch}$, and $O(|\mc{R}|)$ for storing $\Gamma_v$), which is negligible since each element of $\phi_v$, $\phi_{ch}$, and $\Gamma_v$ is a small integer in practice.
On the other hand, implementing the VN update shown by Fig. \ref{fig: VN_update_MIMQBP} for \emph{one} outgoing message has complexity $O(d_v + \lceil \log_2(|\mc{R}|) \rceil)$.
In particular, computing $\Phi_v(l, \mb{s})$ has complexity $O(d_v)$, which allows a binary tree-like parallel implementation; meanwhile, mapping $\Phi_v(l, \mb{s})$ to $R$ based on $\Gamma_v$ has complexity $O(\lceil \log_2(|\mc{R}|) \rceil)$.
The simple implementation for mapping $\Phi_v(l, \mb{s})$ to $R$ also benefits from the use of SDQ rather than the optimal DQ.
Because if an optimal DQ is used instead, it in general requires an additional table of size $|\mc{B}|$ to store this optimal DQ.


\subsection{Remarks}\label{section: remarks at remove}

For each decoding iteration, the design of $Q_{e}: \mc{L} \times \mc{S}^{d_v} \to \mc{X}$ for the MIM-QBP decoding is similar to that of $Q_v$ introduced in Section \ref{section: MIM-QBP decoding at VN}.
In particular, the same RFs $\phi_v$ and $\phi_{ch}$ can be used for the design of $Q_{e}$ and $Q_{v}$ for a given decoding iteration.
We thus ignore the details.

Assume $|\mc{L}|, |\mc{R}|, |\mc{S}|$,  $I_\text{max}$, and $\sigma_d$ are given.
According to the design framework proposed in this section, it is clear that the calculation functions $\Phi_c$ and $\Phi_v$ are fixed, and the TSs $\Gamma_c$ and $\Gamma_v$ are fully determined after choosing RFs $\phi_c$, $\phi_v$, and $\phi_{ch}$.
This results in a great convenience for designing an MIM-QBP decoder, as we only need to find good and even optimal RFs.
Furthermore, we propose an efficient design for RFs in the next section.

\section{An Efficient Design of Reconstruction Functions for MIM-QBP Decoder}
\label{section: design of MIM-QBP decoder}

We illustrated the general framework of the MIM-QBP decoding in Section \ref{section: MIM-QBP decoding}.
The resulting  MIM-QBP decoder works similarly to the quantized BP decoders in \cite{Richardson01capacity, Lee05, Thorpe02}.
In fact, we borrow the terms ``reconstruction function", ``computational domain", ``unreliability", and ``threshold set" from \cite{Richardson01capacity, Lee05, Thorpe02}.
However, the works of \cite{Richardson01capacity, Lee05, Thorpe02} require manual optimization to design the decoder parameters.
On the contrary, we propose an efficient way to systematically design practical MIM-QBP decoders in this section under the general framework of Section \ref{section: MIM-QBP decoding}.
More specifically, we use DE to track the message pmfs, which are given by \eqref{eqn: prob A|X} and \eqref{eqn: P_(S|X) lambda} for the CN update, and \eqref{eqn: prob B|X} and \eqref{eqn: P_(R|X) lambda} for the VN update.
Based on the evolved message pmfs at each iteration, we design the RFs according to the principles discussed in Section \ref{section: MIM-QBP decoding}. We further determine the TSs by the general DP method \cite{he2021dynamic} to obtain the UFs of the MIM-QBP decoder, which aims to maximize $I(X; S)$ for the CN update and $I(X; R)$ for the VN update, respectively.
The detailed design of the UFs is illustrated as follows.


\subsection{Design of $Q_c$ at CN}
\label{section: MIM-QBP decoder at CN}
The design of $Q_c$ for the CN update consists of designing the RF $\phi_c$, the computation of $\mc{A}$ and $P_{A|X}$, and determining the TS $\Gamma_c$, which corresponds to the three subsections under Section \ref{section: MIM-QBP decoding at CN}, respectively.

\subsubsection{Design of $\phi_c$}
We first derive the closed-form of the optimal RF $\phi_c^*$ by setting $\mbb{D} = \mathbb{R}$.
Let $g(r) = P_{X|R}(0|r) - P_{X|R}(1|r)$ for $r \in \mc{R}$ and $g(\mb{r}) = P_{X|\mb{R}}(0 | \mb{r}) - P_{X|\mb{R}}(1 | \mb{r})$ for $\mb{r} \in \mc{R}^{d_c - 1}$.
Note that $P_{X|R}$ can be easily derived from $P_{R|X}$ by Bayes rule with $P_X(0) = P_X(1)=1/2$.
For $r \in \mc{R}$, let
\begin{align}\label{eqn: best phi_c}
    \phi_c^*(r) =
    \begin{cases}
        \mr{sgn}(g(r)) \epsilon & |g(r)| = 1,\\
        -\mr{sgn}(g(r)) \log(|g(r)|) & \text{otherwise},
    \end{cases}
\end{align}
where $\epsilon$ satisfies
\begin{align}\label{eqn: epsilon}
    0 < \epsilon d_c < \min\big\{ &|\log(|g(\mb{r})|) - \log(|g(\mb{r}')|)|:\, \mb{r}, \mb{r}' \in \mc{R}^{d_c - 1}, \nonumber\\
    &g(\mb{r}) \neq g(\mb{r}'), \mr{sgn}(g(\mb{r})) = \mr{sgn}(g(\mb{r}'))  \neq 0 \big\}.
\end{align}
Here $\epsilon$ is a small number close to $0$, which ensures the condition of \eqref{eqn: require phi_c} to be valid for the case of $\phi_c = \phi_c^*$.
Meanwhile, $\epsilon$ is necessary for the following theorem.
\begin{theorem}\label{theorem: phi_c* is optimal}
	Consider the computational domain $\mathbb{D} = \mathbb{R}$ and let $\phi_c = \phi_c^*$. $Q_c$ defined by \eqref{eqn: def of Q_c by Gamma} can maximize $I(X; S)$ among all quantizers from $\mc{R}^{d_c - 1}$ to $\mc{S}$.
\end{theorem}
\begin{IEEEproof}
	See Appendix \ref{appendix: phi_c* is optimal}.
\end{IEEEproof}

Theorem \ref{theorem: phi_c* is optimal} indicates that $\phi_c^*$ is an optimal choice for $\phi_c$ in terms of maximizing $I(X; S)$.
Compared to the function $f(\theta) = \log((e^\theta + 1) / (e^\theta - 1))$ used in \cite{Gallager62} and \cite{Chen05} for the CN update, $\phi_c^*$ is closely related to $f(\theta)$ in the sense that we have $f(|LLR(r)|) = -\log(|g(r)|)$ and $\mr{sgn}(LLR(r)) = \mr{sgn}(g(r))$ for $ r \in \mc{R}$.
This indicates a close connection between the CN update of the BP decoding and that of the MIM-QBP decoding by using the RF $\phi_c = \phi_c^*$.


\begin{corollary}
\label{corollary: phi_c* is optimal}
	Consider the computational domain $\mathbb{D} = \mathbb{R}$ and $\phi_c = \eta \phi_c^*$, where $\eta$ is an arbitrary positive real number.
	$Q_c$ defined by \eqref{eqn: def of Q_c by Gamma} can maximize $I(X; S)$ among all quantizers from $\mc{R}^{d_c - 1}$ to $\mc{S}$.
\end{corollary}

\begin{IEEEproof}
Corollary \ref{corollary: phi_c* is optimal} can be proved similarly to the proof of Theorem \ref{theorem: phi_c* is optimal}.
\end{IEEEproof}
Corollary \ref{corollary: phi_c* is optimal} indicates that for $\mathbb{D} = \mathbb{R}$, the scaled version of $\phi_c^*$  is also optimal for $Q_c$ in the sense of maximizing $I(X; S)$.
This provides a near-optimal solution for designing $\phi_c$ by scaling $\phi_c^*$ for $\mathbb{D} = \mathbb{Z}$.
In the following, we consider $\mathbb{D} = \mathbb{Z}$.
Denote the maximum allowed absolute value of $\phi_c$ by $|\phi_{c}|_{\max}$.
Suppose that there are at most $q_c$ bits allowed for computing the additions by $\Phi_c$, where one bit is needed for representing the sign of each outgoing message.
We have
\begin{equation}\label{eqn: phi_c max}\notag
|\phi_{c}|_\text{max} = \lfloor (2^{q_c-1} - 1)/d_c \rfloor
\end{equation}
such that $\sum_{i = 1}^{d_c} \phi_c(r_i)$ does not overflow.
Let
\[
    |\phi_{c}^*|_\text{max} = \max\{| \phi_c^*(r) |: r \in \mc{R}, g(r) \neq 0\},
\]
where $|\phi_{c}^*|_\text{max} > 0$ holds for a general case.
Inspired by Corollary \ref{corollary: phi_c* is optimal}, we scale $\phi_c^*$ approximately by a factor around $\eta_c = |\phi_{c}|_\text{max} / |\phi_{c}^*|_\text{max}$ to obtain $\phi_c: \mc{R} \to \mathbb{Z}$ given by
\begin{equation}\label{eqn: design of phi_c}
    \phi_c(r) =
    \begin{cases}
         \mr{sgn}(g(r)) \max\{1, \mr{round}(\eta_c |\phi_c^*(r)|) \}
         & g(r) \neq 0,\\
        |\phi_{c}|_{\max} & g(r) = 0,
    \end{cases}
\end{equation}
where $\rm{round}()$ returns the closest integer of a floating number.
In this case, we have $|\phi_c(r)| \le |\phi_{c}|_\text{max}, \forall\, r \in \mc{R}$.
Note that for $g(r) \neq 0$, we make $|\phi_c(r)| \geq 1$ to ensure $\mr{sgn}(\phi_c(r)) = \mr{sgn}(g(r))$.
In particular, for $|g(r)| = 1$, we have $|\phi_c(r)| = 1$ since $\epsilon$ is close to $0$.
Moreover, for $g(r) = 0$, $r$ has the most unreliability since $P_{X|R}(0|r)=P_{X|R}(1|r)=1/2$.
In this case, we use $\phi_c(r) = |\phi_{c}|_\text{max}$ according to the design principle \eqref{eqn: require phi_c}.

\subsubsection{Computation of $\mc{A}$ and $P_{A|X}$}
By using the RF $\phi_c$ given by \eqref{eqn: design of phi_c}, the function $\Phi_c$ defined by \eqref{eqn: def of Phi_c} is capable of mapping $\mc{R}^{d_c - 1}$ to $\mathbb{Z}$ and obtaining the C2V computational messages represented by $q_c$-bit integers.
Meanwhile, the integer set $\mc{A}$ and the message pmf $P_{A|X}$ need to be computed for the decoder design in the quantization step.
Note that directly adopting the computation in the ways of \eqref{eqn: def of mc_A} and \eqref{eqn: prob A|X} can be a prohibitive task when $|\mc{R}|^{d_c-1}$ is large.
In the following, we propose a fast method to compute $\mc{A}$ and $P_{A|X}$.
The essential idea is using DP \cite[Section 15.3]{IntroAlgo01} to handle the $d_c-1$ V2C message pmfs one by one, and carry out the computation with the previous intermediate results and the current V2C message pmf.

Starting from \eqref{eqn: joint P_R|X}, for $k \geq 1$ and $\mb{R} \in \mc{R}^k$, we have
\begin{equation*}
    P_{\mb{R}|X}(\mb{r}|x) = \left(\frac{1}{2}\right)^{\dim(\mb{r}) - 1} \sum_{\mb{x}: \oplus \mb{x} = x} \prod_{i = 1}^{\dim(\mb{r})} P_{R|X}(r_i|x_i),
\end{equation*}
which denotes the joint distribution of $k$ V2C messages conditioned on the channel input bit.
Denote the set of distinct values of $\Phi_c(\mb{r}), \forall \, \mb{r} \in \mc{R}^{k}$ by
\begin{equation*}
    \mc{A}_k = \{a_{k, 1}, a_{k, 2}, \ldots, a_{k, |\mc{A}_k|}\}.
\end{equation*}
Let $A_k$ be a random variable which takes values from $\mc{A}_k$.
Motivated by \eqref{eqn: P(R|0) - P(R|1)}, we define $\delta_k^{+}(\cdot)$ and $\delta_k^{-}(\cdot)$ as
\begin{align*}
    &\delta_k^{\pm}(a_{k, i})\\
    =& \sum_{\mb{r} \in \mc{R}^{k}, \Phi_c(\mb{r}) = a_{k,i}} (P_{\mb{R} | X}(\mb{r} | 0) \pm P_{\mb{R} | X}(\mb{r} | 1))\\
    =& P_{A_k|X}(a_{k, i}|0) \pm P_{A_k|X}(a_{k, i}|1).
\end{align*}
Based on $\Phi_c$ given by \eqref{eqn: def of Phi_c}, for $\alpha, \beta \in \mathbb{R}$, we define $\diamond$ as a binary operator such that
\[
    \alpha \diamond \beta = \mr{sgn}(\alpha)\mr{sgn}(\beta)(|\alpha| + |\beta|).
\]
Then we have the following Proposition \ref{proposition: P(A_k|0) - P(A_k|1)}.
\begin{proposition}\label{proposition: P(A_k|0) - P(A_k|1)}
For $k = 1$, we have
\begin{align}\label{eqn: P(A1|0) - P(A1|1)}
    \mc{A}_k &= \{\phi_c(r):\, r \in \mc{R} \}, \text{~and~}\nonumber\\
    \delta_k^{\pm}(a_{k, i}) &= \sum_{r \in \mc{R}, \phi_c(r) = a_{k, i}} (P_{R|X}(r|0) \pm P_{R|X}(r|1)).
\end{align}
For $k > 1$, we have
\begin{align}\label{eqn: P(A_k|0) - P(A_k|1)}
    &\quad\quad\,\mc{A}_k = \{\phi_c(r) \diamond a_{k-1, j}:\, r \in \mc{R}, a_{k-1, j} \in \mc{A}_{k-1} \},\text{~and~}\nonumber\\[-0.5em]
    &\delta_k^{\pm}(a_{k, i}) = \nonumber\\[-0.5em]
    &\sum_{\substack{r \in \mc{R}, a_{k-1, j} \in \mc{A}_{k-1},\\ \phi_c(r) \diamond a_{k-1, j} = a_{k, i}}} \frac{1}{2} (P_{R|X}(r|0) \pm P_{R|X}(r|1))  \delta_{k-1}^{\pm}(a_{k-1, j}).
\end{align}
\end{proposition}
\begin{IEEEproof}
For $k = 1$, \eqref{eqn: P(A1|0) - P(A1|1)} holds obviously.
For $k > 1$, we have
\begin{align*}
    \mc{A}_k &= \{\Phi_c(\mb{r}) : \mb{r} \in \mc{R}^{k}\}\\[-0.5em]
    &= \{\phi_c(r) \diamond \Phi_c(\mb{r}):\, r \in \mc{R}, \mb{r} \in \mc{R}^{k-1} \}\\[-0.5em]
    &= \{\phi_c(r) \diamond a_{k-1, j}:\, r \in \mc{R}, a_{k-1, j} \in \mc{A}_{k-1} \};
\end{align*}
meanwhile, we have
\begin{align*}
    &\delta_k^{\pm}(a_{k, i})\nonumber\\[-0.2em]
    =& \sum_{\mb{r} \in \mc{R}^{k}, \Phi_c(\mb{r}) = a_{k,i}} (P_{\mb{R} | X}(\mb{r} | 0) \pm P_{\mb{R} | X}(\mb{r} | 1))\nonumber\\
    =& \sum_{\mb{r} \in \mc{R}^{k}, \Phi_c(\mb{r}) = a_{k,i}} \left(\frac{1}{2}\right)^{k-1} \prod_{u = 1}^{k} (P_{R | X}(r_u | 0) \pm P_{R | X}(r_u | 1))\nonumber\\[-0.2em]
    =& \sum_{r_k \in \mc{R}} \frac{1}{2} (P_{R|X}(r_k|0) \pm P_{R|X}(r_k|1)) \times \\[-0.2em]
    &\sum_{\substack{\mb{r} \in \mc{R}^{k-1},\\ \phi_c(r) \diamond \Phi_c(\mb{r}) = a_{k,i}}} \left(\frac{1}{2}\right)^{k-2} \prod_{u = 1}^{k-1} (P_{R | X}(r_u | 0) \pm P_{R | X}(r_u | 1))\nonumber\\[-0.2em]
    =& \sum_{r_k \in \mc{R}} \frac{1}{2} (P_{R|X}(r_k|0) \pm P_{R|X}(r_k|1)) \times \\[-0.7em]
    &\quad\quad\quad\quad \sum_{\substack{\mb{r} \in \mc{R}^{k-1},\\ \phi_c(r) \diamond \Phi_c(\mb{r}) = a_{k,i}}} (P_{\mb{R} | X}(\mb{r} | 0) \pm P_{\mb{R} | X}(\mb{r} | 1))\nonumber\\[-0.5em]
    =& \sum_{r \in \mc{R}} \frac{1}{2} (P_{R|X}(r|0) \pm P_{R|X}(r|1)) \!\!\!\!\! \sum_{\substack{a_{k-1, j} \in \mc{A}_{k-1},\\ \phi_c(r) \diamond a_{k-1, j} = a_{k, i}}} \!\!\!\!\! \delta_{k-1}^{\pm}(a_{k-1, j}).
\end{align*}
This completes the proof.
\end{IEEEproof}

Proposition \ref{proposition: P(A_k|0) - P(A_k|1)} sets up the recursive computation rules, where for $k > 1$, $\mc{A}_{k-1}$ and $\delta_{k-1}^{\pm}$ are the previous intermediate computation results, and they are used to compute $\mc{A}_k$ and $\delta_{k}^{\pm}$ together with the $k$-th V2C message $R$ and its pmf $P_{R|X}$.
According to Proposition \ref{proposition: P(A_k|0) - P(A_k|1)}, we can compute $\mc{A}_1$, $\delta_{1}^{\pm}$, $\mc{A}_2$,  $\delta_{2}^{\pm}$, $\ldots$, $\mc{A}_{d_c-1}$, $\delta_{d_c-1}^{\pm}$ sequentially, and obtain the message pmf $P_{A_{k}|X}$ by
\begin{align}
\notag
\left\{
\begin{array}{l}
P_{A_{k}|X}(a_{{k}, i} | 0) = (\delta_{{k}}^{+}(a_{{k}, i}) + \delta_{{k}}^{-}(a_{{k}, i}) ) / 2,\\
P_{A_{k}|X}(a_{{k}, i} | 1) = (\delta_{{k}}^{+}(a_{{k}, i}) - \delta_{{k}}^{-}(a_{{k}, i}) ) / 2.
\end{array}
\right.
\end{align}
Then, the integer set $\mc{A}$ and the message pmf $P_{A|X}$ are equal to $\mc{A}_{d_c-1}$ and $P_{A_{d_c-1}|X}$, respectively.
We summarize the corresponding computation by Algorithm \ref{algo: compute P_A|X}.
Since $|\mc{A}_{k-1}|$ in line \ref{code: comp P(A|X)} of Algorithm \ref{algo: compute P_A|X} is upper-bounded by $2^{q_c}$, the complexity of Algorithm \ref{algo: compute P_A|X} is $O(d_c 2^{q_c} |\mc{R}|)$.

\begin{table}[t!]
\begin{algorithm}[H]
\caption{Computation of $\mc{A}$ and $P_{A|X}$}
\label{algo: compute P_A|X}
\begin{algorithmic}[1]
\setstretch{1.3}
\REQUIRE $P_{R|X}, \phi_c, d_c$.
\ENSURE $\mc{A}$ and $P_{A|X}$.

\STATE  Set $\mc{A}_k = \emptyset$ and $\delta_{k}^{\pm}(\cdot) = 0$ for $k = 1, 2, \ldots, d_c - 1$.

\FOR    {$r \in \mc{R}$}
    \STATE  $\mc{A}_1 = \mc{A}_1 \cup \{\phi_c(r)\}$. $//$\textit{See \eqref{eqn: P(A1|0) - P(A1|1)}}
    \STATE  $\delta_{1}^{\pm}(\phi_c(r)) \,{+\!\!=}\, P_{R|X}(r|0) \pm P_{R|X}(r|1)$.
\ENDFOR

\FOR    {$k = 2, 3, \ldots, d_c - 1$}
    \FOR    {$r \in \mc{R}, a_{k-1, j} \in \mc{A}_{k-1}$}\label{code: comp P(A|X)}
        \STATE  $a_{k, i} = \phi_c(r) \diamond a_{k-1, j}$.
        \STATE  $\mc{A}_k = \mc{A}_k \cup \{a_{k, i}\}$. $//$\textit{See \eqref{eqn: P(A_k|0) - P(A_k|1)}}
        \STATE  $\delta_{k}^{\pm}(a_{k, i}){+\!\!=} \frac{1}{2} (P_{R|X}(r|0) \pm P_{R|X}(r|1)) \delta_{k-1}^{\pm}(a_{k-1, j})$.
    \ENDFOR
\ENDFOR

\FOR    {$k = 1, 2, \ldots, d_c - 1$}
    \FOR    {$a_{k, i} \in \mc{A}_{k}$}
        \STATE  $P_{A_k|X}(a_{k, i} | 0) = (\delta_{k}^{+}(a_{k, i}) + \delta_{k}^{-}(a_{k, i}) ) / 2$.
        \STATE $P_{A_k|X}(a_{k, i} | 1) = (\delta_{k}^{+}(a_{k, i}) - \delta_{k}^{-}(a_{k, i}) ) / 2$.
    \ENDFOR
\ENDFOR

\STATE  $\mc{A} = \mc{A}_{d_c-1}$.
\STATE  $P_{A|X} = P_{A_{d_c - 1}|X}$.

\RETURN $\mc{A}$ and $P_{A|X}$.
\end{algorithmic}
\end{algorithm}
\end{table}

\subsubsection{Determining $\Gamma_c$}
Based on $\mc{A}$ and $P_{A|X}$, we can compute the optimal SDQ $\Lambda_c$ by \eqref{eqn: def of Lambda_c}.
Then we can obtain the pmf $P_{S|X}$ and the TS $\Gamma_c$ given by \eqref{eqn: P_(S|X) lambda} and \eqref{eqn: def of Gamma_c}, respectively.
The UF $Q_c$ operates as \eqref{eqn: def of Q_c by Gamma} to compute the C2V messages, where we only need to store $\phi_c$ and $\Gamma_c$ for the CN update.
The computational complexity is $O(d_c + d_c \lceil \log_2(|\mc{S}|)\rceil)$ for one CN per iteration (including integer comparisons and additions with bit width $q_c$).

\subsection{Design of $Q_v$ at VN}
\label{section: MIM-QBP decoder at VN}
The design of $Q_v$ for the VN update contains designing the RFs $\phi_v$ and $\phi_{ch}$, the computation of $\mc{B}$ and $P_{B|X}$, and determining the TS $\Gamma_v$, which corresponds to the three subsections under Section \ref{section: MIM-QBP decoding at VN}, respectively.
\subsubsection{Design of $\phi_v$ and $\phi_{ch}$}
We first derive the closed-form of the optimal RFs $\phi_v^*$ and $\phi_{ch}^*$ by setting $\mbb{D} = \mathbb{R}$.
For $s \in \mc{S}$ and $l \in \mc{L}$, let
\begin{align}\label{eqn: best phi_v}
\left\{
    \begin{array}{l}
    \phi_v^*(s) = \log ({P_{S|X}(s|0) }/{ P_{S|X}(s|1)}),\\
    \phi_{ch}^*(l) = \log ({P_{L|X}(l|0) }/{ P_{L|X}(l|1)}).
    \end{array}
\right.
\end{align}
We can easily verify that the condition of \eqref{eqn: require phi_v} holds for $\phi_v = \phi_v^*$ and $\phi_{ch} = \phi_{ch}^*$.
Moreover, $(\phi_v^*, \phi_{ch}^*)$ is proved to be an optimal choice for $(\phi_c, \phi_{ch})$ in terms of maximizing $I(X; R)$ by the following theorem.
\begin{theorem}
\label{theorem: phi_v* is optimal}
	Consider the computational domain $\mathbb{D} = \mathbb{R}$.
	Let $\phi_v = \phi_v^*$ and $\phi_{ch} = \phi_{ch}^*$.
	$Q_v$ defined by \eqref{eqn: def of Q_v by Gamma} can maximize $I(X; R)$ among all quantizers from $\mc{L} \times \mc{S}^{d_v - 1}$ to $\mc{R}$.
\end{theorem}

\begin{IEEEproof}
See Appendix \ref{appendix: phi_v* is optimal}.
\end{IEEEproof}

Note that we have $\phi_v^*(s) = LLR(s) + \log(P_X(1) / P_X(0))$ and $\phi_{ch}^*(l) = LLR(l) + \log(P_X(1) / P_X(0))$.
With $P_X(1) = P_X(0) = 1/2$, $\phi_v^*$ and $\phi_{ch}^*$ implies a close relation between the VN update of the BP decoding and that of the MIM-QBP decoding by using $\phi_v = \phi_v^*$ and $\phi_{ch} = \phi_{ch}^*$.
To realize fixed-point implementation, we further consider the following corollary.
\begin{corollary}
\label{corollary: phi_v* is optimal}
	Consider $\mathbb{D} = \mathbb{R}$.
	Let $\phi_v = \eta \phi_v^*$ and $\phi_{ch} = \eta \phi_{ch}^*$, where $\eta$ is an arbitrary positive real number.
	$Q_v$ defined by \eqref{eqn: def of Q_v by Gamma} can maximize $I(X; R)$ among all quantizers from $\mc{L} \times \mc{S}^{d_v - 1}$ to $\mc{R}$.
\end{corollary}

\begin{IEEEproof}
Corollary \ref{corollary: phi_v* is optimal} can be proved in a way similar to the proof of Theorem \ref{theorem: phi_v* is optimal}.
\end{IEEEproof}

Corollary \ref{corollary: phi_v* is optimal} indicates that  for $\mathbb{D} = \mathbb{R}$, the scaled versions of $\phi_v^*$ and $\phi_{ch}^*$ are also optimal for $Q_v$ with respect to maximizing $I(X; R)$.
This provides a near-optimal solution for designing $\phi_v$ and $\phi_{ch}$ by scaling $\phi_v^*$ and $\phi_{ch}^*$, respectively, for $\mathbb{D} = \mathbb{Z}$.
In the following, we consider $\mathbb{D} = \mathbb{Z}$.
\color{black}
Denote the maximum allowed absolute value of $\phi_v(\cdot)$ and $\phi_{ch}(\cdot)$ by $|\phi_{v,ch}|_\text{max}$.
Suppose that there are at most $q_v$ bits allowed for computing the additions by $\Phi_v$.
Then, $|\phi_{v,ch}|_\text{max}$ can be taken as
\begin{equation}\label{eqn: phi_v max}
\notag
|\phi_{v,ch}|_\text{max} = \lfloor (2^{q_v-1} - 1)/(d_v + 1) \rfloor
\end{equation}
such that $\phi_{ch}(l) + \sum_{i = 1}^{d_v} \phi_v(s_i)$ does not overflow.
Let
\[
|\phi_{v, ch}^*|_\text{max} = \max ( \{| \phi_v^*(s) |: s \in \mc{S}\} \cup \{| \phi_{ch}^*(l) |: l \in \mc{L}\} ).
\]
Note that $|\phi_{v, ch}^*|_\text{max} > 0$ holds for a general case.
Then, inspired by Corollary \ref{corollary: phi_v* is optimal}, we respectively scale $\phi_v^*$ and $\phi_{ch}^*$ by a factor around $\eta_v = |\phi_{v, ch}|_\text{max} / |\phi_{v, ch}^*|_\text{max}$ to obtain $\phi_v: \mc{S} \to \mbb{Z}$ and $\phi_{ch}: \mc{L} \to \mbb{Z}$ as follows
\begin{align}\label{eqn: design of phi_v}
\left\{
\begin{array}{l}
\phi_v(s) = \mr{sgn}(\phi_v^*(s)) \mr{round}( \eta_v |\phi_v^*(s)|),\\
\phi_{ch}(l) = \mr{sgn}(\phi_{ch}^*(l)) \mr{round}( \eta_v |\phi_{ch}^*(l)| ).
\end{array}
\right.
\end{align}	
In this case, we have $|\phi_v(s)| \le |\phi_{v, ch}|_\text{max}$ and $|\phi_{ch}(l)| \le |\phi_{v, ch}|_\text{max}$.

\subsubsection{Computation of $\mc{B}$ and $P_{B|X}$}
By using the RFs given by \eqref{eqn: design of phi_v}, the function $\Phi_v$ defined by \eqref{eqn: def of Phi_v} is able to map $\mc{L} \times \mc{S}^{d_v - 1}$ to $\mathbb{Z}$ and obtain the V2C computational messages represented by $q_v$-bit integers.
Similar to the case of the CN update, the integer set $\mc{B}$ and the message pmf $P_{B|X}$ need to be computed accordingly.
However, directly computing $\mc{B}$ and $P_{B|X}$ by \eqref{eqn: def of mc_B} and \eqref{eqn: prob B|X} requires large complexity if $|\mc{L}| \times |\mc{S}|^{d_v-1}$ is large.
In the following, we propose a fast method to compute $\mc{B}$ and $P_{B|X}$.
Similar to the case of calculating $\mc{A}$ and $P_{A|X}$, the essential idea is adopting DP \cite[Section 15.3]{IntroAlgo01} to handle the $d_v$ message  pmfs (one channel-to-variable message pmf and $d_v-1$ C2V message pmfs) one by one,
and carry out the computation with the previous intermediate results and the incoming current  message pmf.

Beginning with \eqref{eqn: joint P_L,S|X}, for $k \geq 0$, $L \in \mc{L}$, and $\mb{S} \in \mc{S}^k$, we have
\begin{equation*}
P_{L,\mb{S}|X}(l, \mb{s}|x) =  P_{L|X}(l|x) \prod_{i = 1}^{\dim(\mb{s})} P_{S|X}(s_i|x),
\end{equation*}
which denotes the joint distribution of incoming message $(L, \mb{S}) \in \mc{L} \times \mc{S}^{k}$ conditioned on the channel input bit $X$ at a VN.
Denote the set of distinct values of $\Phi_v(l, \mb{s}), \forall \, l \in \mc{L}, \forall \, \mb{s} \in \mc{S}^{k}$ by
\begin{equation*}
\mc{B}_k = \{b_{k, 1}, b_{k, 2}, \ldots, b_{k, |\mc{B}_k|}\},
\end{equation*}
In particular, for $k = 0$, we have
\begin{equation*}
P_{L,\mb{S}|X}(l, \mb{s}|x) =  P_{L|X}(l|x), ~\text{and}~\Phi_v(l, \mb{s}) = \phi_{ch}(l).
\end{equation*}
Let $B_k$ be a random variable which takes values from $\mc{B}_k$.
Then we have the following Proposition \ref{proposition: P(B_k|X)}.

\begin{proposition}\label{proposition: P(B_k|X)}
For $k = 0$, we have
\begin{align}\label{eqn: P(B0|X)}
    \mc{B}_k &= \{\phi_{ch}(l):\, l \in \mc{L}\},\text{~and~}\nonumber\\[-0.5em]
    P_{B_k|X}(b_{k, i}|x) &= \sum_{l \in \mc{L}, \phi_{ch}(l) = b_{k, i}} P_{L|X}(l|x).
\end{align}
For $k > 0$, we have
\begin{align}\label{eqn: P(B_k|X)}
    &\mc{B}_k = \{\phi_v(s) + b_{k-1, j}:\, s \in \mc{S}, b_{k-1, j} \in \mc{B}_{k-1}\},\text{~and~}\nonumber\\[-0.5em]
    &P_{B_k|X}(b_{k, i}|x) = \!\!\!\!\!\! \sum_{\substack{s \in \mc{S}, b_{k-1, j} \in \mc{B}_{k-1},\\\phi_v(s) + b_{k-1, j} = b_{k, i}}} \!\!\!\!\!\! P_{S|X}(s|x) P_{B_{k-1}|X}(b_{k-1, j}|x).
\end{align}
\end{proposition}

\begin{IEEEproof}
For $k = 0$, \eqref{eqn: P(B0|X)} holds obviously.
For $k > 0$, we have
\begin{align*}
    \mc{B}_k &= \{\Phi_v(l, \mb{s}) : l \in \mc{L}, \mb{s} \in \mc{S}^{k}\}\\[-0.5em]
    &= \{\phi_v(s) + \Phi_v(l, \mb{s}) : s \in \mc{S}, l \in \mc{L}, \mb{s} \in \mc{S}^{k-1}\}\\[-0.5em]
    &= \{\phi_v(s) + b_{k-1, j}:\, s \in \mc{S}, b_{k-1, j} \in \mc{B}_{k-1}\};
\end{align*}
meanwhile, we have
\begin{align*}
    &P_{B_k|X}(b_{k, i}|x) \\
    =& \sum_{l \in \mc{L}, \mb{s} \in \mc{S}^{k}, \Phi_v(l, \mb{s}) = b_{k,i}} P_{L, \mb{S} | X}(l, \mb{s} | x)\nonumber\\[-0.3em]
    =& \sum_{l \in \mc{L}, \mb{s} \in \mc{S}^{k}, \Phi_v(l, \mb{s}) = b_{k,i}} P_{L|X}(l|x) \prod_{i=1}^{k} P_{S|X}(s_i|x)\nonumber\\[-0.3em]
    =& \sum_{s_k \in \mc{S}} P_{S|X}(s_k|x) \!\!\!\! \sum_{\substack{l \in \mc{L}, \mb{s} \in \mc{S}^{k-1},\\ \phi_v(s_k) + \Phi_v(l, \mb{s}) = b_{k,i}}} \!\!\!\!\!\!\!\!\!\!  P_{L|X}(l|x) \prod_{i=1}^{k-1} P_{S|X}(s_i|x)\nonumber\\[-0.3em]
    =& \sum_{s_k \in \mc{S}} P_{S|X}(s_k|x) \!\!\!\! \sum_{\substack{l \in \mc{L}, \mb{s} \in \mc{S}^{k-1},\\ \phi_v(s_k) + \Phi_v(l, \mb{s}) = b_{k,i}}} P_{L, \mb{S} | X}(l, \mb{s} | x)\nonumber\\[-0.3em]
    =& \sum_{s \in \mc{S}} P_{S|X}(s|x) \sum_{\substack{b_{k-1, j} \in \mc{B}_{k-1},\\ \phi_v(s) + b_{k-1, j} = b_{k, i}}} P_{B_{k-1}|X}(b_{k-1, j}|x).
\end{align*}
This completes the proof.
\end{IEEEproof}

Proposition \ref{proposition: P(B_k|X)} sets up the recursive computation rules, where we use $\mc{B}_{k-1}$ and $b_{k-1}$ as the previous intermediate computation results for $k > 0$ to compute $\mc{B}_k$ and $b_k$ together with the $k$-th incoming message $S$ and its pmf $P_{S|X}$.
According to Proposition \ref{proposition: P(B_k|X)}, we can compute $\mc{B}_0$, $P_{B_0|X}$, $\mc{B}_1$, $P_{B_1|X}$, $\ldots$, $\mc{B}_{d_v-1}$, $P_{B_{d_v-1}|X}$ sequentially.
Then, the integer set $\mc{B}$ and the message pmf $P_{B|X}$ are equal to $\mc{B}_{d_v-1}$ and $P_{B_{d_v-1}|X}$, respectively.
We summarize the corresponding computation by Algorithm \ref{algo: compute P_B|X}.
Since $|\mc{B}_{k-1}|$ in line \ref{code: comp P(B|X)} of Algorithm \ref{algo: compute P_B|X} is upper-bounded by $2^{q_v}$, the complexity of Algorithm \ref{algo: compute P_B|X} is $O(d_v 2^{q_v} |\mc{S}|)$.
\begin{table}[t!]
\begin{algorithm}[H]
\caption{Computation of $\mc{B}$ and $P_{B|X}$}
\label{algo: compute P_B|X}
\begin{algorithmic}[1]
\setstretch{1.3}
\REQUIRE $\phi_v, \phi_{ch}, P_{S|X}, P_{L|X}, d_v$.
\ENSURE $\mc{B}$ and $P_{B|X}$.

\STATE  Set $\mc{B}_k = \emptyset$ and $P_{B_k|X}(\cdot|x) = 0$ for $k = 0, 1, \ldots, d_v - 1$ and for $x = 0, 1$.

\FOR    {$l \in \mc{L}$}
    \STATE  $\mc{B}_0 = \mc{B}_0 \cup \{\phi_{ch}(l)\}$. $//$\textit{See \eqref{eqn: P(B0|X)}}
    \STATE  $P_{B_0|X}(\phi_{ch}(l) | x) \,{+\!\!=}\, P_{L|X}(l|x)$ for $x = 0, 1$.
\ENDFOR

\FOR    {$k = 1, 2, \ldots, d_v - 1$}
    \FOR    {$s \in \mc{S}, b_{k-1, j} \in \mc{B}_{k-1}$}\label{code: comp P(B|X)}
        \STATE  $b_{k, i} = \phi_v(s) + b_{k-1, j}$.
        \STATE  $\mc{B}_k = \mc{B}_k \cup \{b_{k, i}\}$. $//$\textit{See \eqref{eqn: P(B_k|X)}}
        \STATE  $P_{B_k|X}(b_{k, i}|x) \,{+\!\!=}\, P_{S|X}(s|x) P_{B_{k-1}|X}(b_{k-1, j}|x))$ for $x = 0, 1$.
    \ENDFOR
\ENDFOR

\STATE  $P_{B|X} = P_{B_{d_v - 1}|X}$.

\RETURN $\mc{B}$ and $P_{B|X}$.
\end{algorithmic}
\end{algorithm}
\end{table}

\subsubsection{Determining $\Gamma_v$}
Based on $\mc{B}$ and $P_{B|X}$, we can compute the optimal SDQ $\Lambda_v$ by \eqref{eqn: def of Lambda_v}.
Then we obtain the pmf $P_{R|X}$ and the TS $\Gamma_v$ given by \eqref{eqn: P_(R|X) lambda} and \eqref{eqn: def of Gamma_v}, respectively.
The UF $Q_v$ operates as \eqref{eqn: def of Q_v by Gamma} to compute the messages passed from VN to CN, where we need to store $\phi_v$, $\phi_{ch}$, and $\Gamma_v$ for the VN update.
The computational complexity is $O(d_v + d_v \lceil \log_2(|\mc{R}|)\rceil)$ for one VN per iteration (including integer comparisons and additions with bit width $q_v$).
\subsection{Remarks}\label{section: remarks at design}

As illustrated by Section \ref{section: remarks at remove}, the design of $Q_{e}$ is quite similar to that of $Q_{v}$.
In particular, the same RFs $\phi_v$ and $\phi_{ch}$ can be used for the design of $Q_{e}$ and $Q_{v}$ for a given decoding iteration, and the design of $Q_{e}$ has a complexity of $O(d_v 2^{q_v} |\mc{S}| + 2^{2 q_v} |\mc{X}|)$ (for one decoding iteration).
After the design of $Q_e$, implementing $Q_e$ for one VN for one iteration during decoding has complexity $O(d_v)$, which is equal to the complexity of addition operations with $q_v$ bit widths.

After setting desirable maximum allowed bit widths $q_c$ for the CN update and $q_v$ for the VN update, we can obtain the near-optimal practical RFs $\phi_c$ by \eqref{eqn: design of phi_c} and $(\phi_v, \phi_{ch})$ by \eqref{eqn: design of phi_v}, which are scaled versions of their optimal counterparts $\phi_{c}^*$, $\phi_{v}^*$, and $\phi_{ch}^*$, respectively.
We remark that if $q_c$ and $q_v$ are sufficiently large (e.g., infinite precision is allowed), $\phi_{c}$, $\phi_{v}$, and $\phi_{ch}$ coincide with their optimal counterparts, which can make the MIM-QBP decoder always maximize the MI between the outgoing message and the coded bit according to Corollaries \ref{corollary: phi_c* is optimal} and \ref{corollary: phi_v* is optimal}.
In this case, the MIM-QBP decoder works the same as the MIM-LUT decoder without table decomposition.
On the other hand, for practical implementations with low precision, our design of the RFs based on \eqref{eqn: design of phi_c} and \eqref{eqn: design of phi_v} may not be optimal anymore.
This is because the scaling becomes less accurate with the decrease of $q_c$ and $q_v$.
However, our design can still achieve good error rate performance according to extensive simulation results.
Furthermore, we believe that the only way to obtain the optimal RFs with low precision is to conduct the exhaustive search for $\phi_{c}$, $\phi_{v}$, and $\phi_{ch}$, which will be tedious and time-consuming.

At this point, it is clear that setting proper $(|\mc{L}|, |\mc{R}|, |\mc{S}|, I_\text{max}, q_c, q_v, \sigma_d)$ is sufficient for our systematic design of the RCQ parameters of an MIM-QBP decoder.
Moreover, we can set $(|\mc{L}|, |\mc{R}|, |\mc{S}|, I_\text{max}, q_c, q_v)$ according to practical considerations, e.g., complexity.
On the other hand, similar to the MIM-LUT decoder, the performance of the MIM-QBP decoder also depends greatly on the choice of $\sigma_d$.
We also refer to $\sigma^*$ given by \eqref{sigma} as the design threshold of the MIM-QBP decoder.
Our simulation results show that the MIM-QBP decoder designed at $\sigma_d = \sigma^*$ can achieve desirable error rate performance across a wide SNR range.

\section{Simulation Results}\label{section: simulation results}

In this section, we evaluate the error rate performance of the proposed MIM-QBP decoders via Monte-Carlo simulations.
Assume binary phase-shift keying (BPSK) transmission over the AWGN channel.
We design the MIM-QBP decoder by fixing $|\mc{L}| = |\mc{R}| = |\mc{S}| = 8/16$ (3-/4-bit decoder) for all iterations.
We specify $q_c$ (bit width used for CN update), $q_v$ (bit width used for VN update), and $\sigma_d$ (design noise standard deviation) for each specific example.
At least 100 frame errors are collected for each simulated SNR.

\begin{figure}[!h]
	\centering
	\subfigure[A maximum of 10 iterations.]{%
		\includegraphics[scale = 0.65]{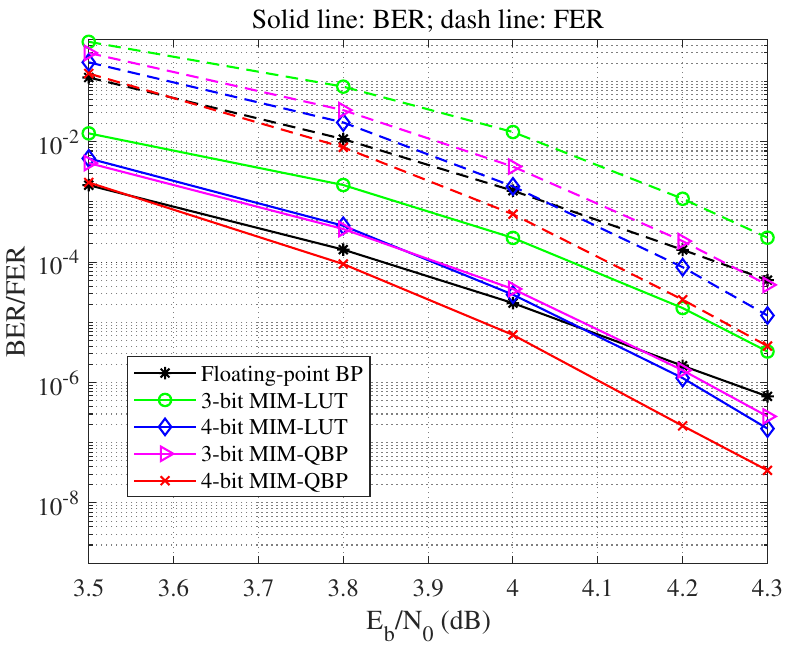}}
	\subfigure[A maximum of 30 iterations.]{%
		\includegraphics[scale = 0.65]{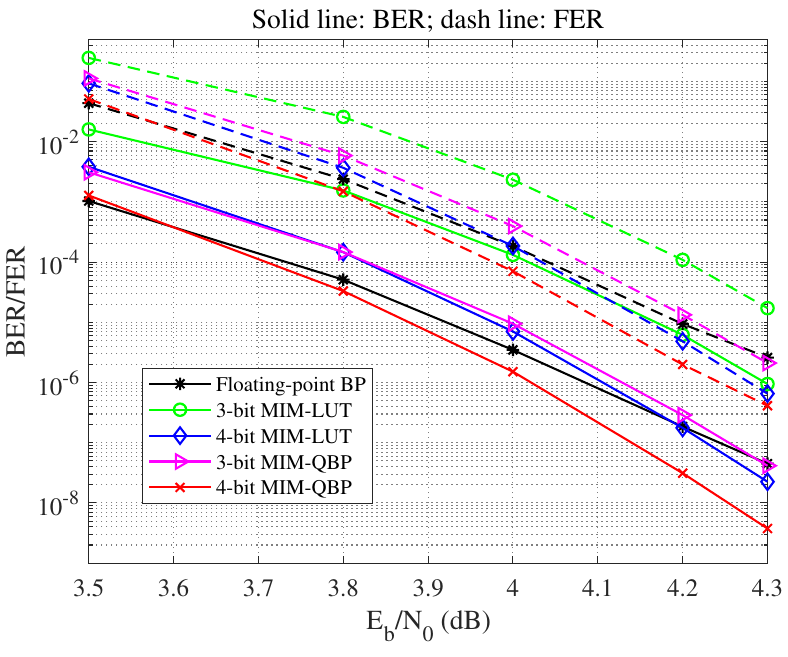}
	}
	\caption{BER and FER simulation results for the (6, 32) code \cite{IEEESTD06} of length 2048 and rate 0.84. We set $(q_c, q_v) = (10, 8)$ and $(q_c, q_v) = (12, 10)$ for the 3-bit and 4-bit MIM-QBP decoders, respectively.}
	\label{fig: BER_2048_6_32}
\end{figure}

\begin{example}\label{eg: code 2048}
Consider the regular (6, 32) LDPC code taken from
\cite{IEEESTD06}.
This code has length 2048 and rate 0.84.
We use $(q_c, q_v) = (10, 8)/(12, 10)$ to design the 3-/4-bit MIM-QBP decoder at $\sigma_d = 0.5343/0.5417$, respectively.
The bit error rate (BER) and frame error rate (FER) of different decoders are illustrated by Fig. \ref{fig: BER_2048_6_32}.
We observe that our proposed 4-bit MIM-QBP decoder outperforms both the 4-bit MIM-LUT decoder \cite{meidlinger2020design} and the floating-point BP decoder, with 10/30 iterations, respectively.
Moreover, the 3-bit MIM-QBP decoder outperforms the 3-bit MIM-LUT decoder by about $0.15$ dB, and even surpasses the floating-point BP decoder at the high SNR region.
Note that the degradation of the performance of the floating-point BP decoder is mainly caused by the ($8, 8$) trapping sets in the code graph \cite{ryan2009channel}.
In contrast, our proposed MIM-QBP decoders are designed based on maximizing the mutual information, which may improve the ability to overcome certain trapping sets during the decoding process \cite{Romero16}.

\end{example}

\begin{example}
Consider the regular (3, 6) LDPC code with identifier 8000.4000.3.483 taken from \cite{MacKay}.
This code has length 8000 and rate 0.5.
We use $(q_c, q_v) = (9, 8)/(10, 10)$ to design the 3-/4-bit MIM-QBP decoder at $\sigma_d = 0.8479/0.8660$, respectively.
The BER performance of different decoders is presented by Fig. \ref{fig: BER_8000_3_6}.
Note that the 3-/4-bit non-uniform QBP decoder \cite{	Lee05} requires $(q_c, q_v) = (9, 8)/(12, 10)$, respectively.
In addition, the design of the corresponding decoders involves much manual optimization, while our proposed MIM-QBP decoders are designed systematically.

\begin{figure}[h]
	\centering
	\includegraphics[scale = 0.65]{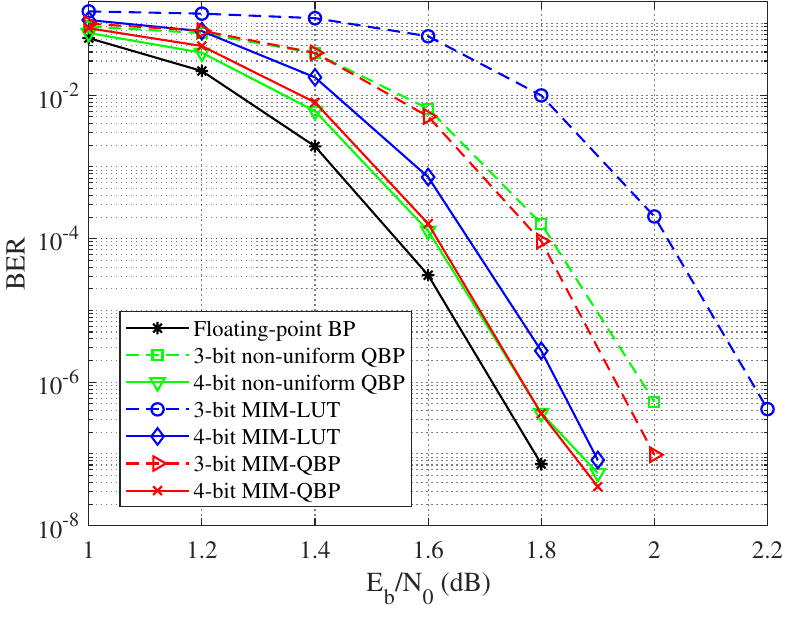}
	\caption{BER simulation results for the (3, 6) code (with identifier 8000.4000.3.483 in  \cite{MacKay}) of length 8000 and rate 0.5. A maximum of 50 iterations is used. We set $(q_c, q_v) = (9, 8)$ and $(q_c, q_v) = (10, 10)$ for the 3-bit and 4-bit MIM-QBP decoders, respectively.}
	\label{fig: BER_8000_3_6}
\end{figure}

From Fig. \ref{fig: BER_8000_3_6}, we observe that the 3-bit MIM-QBP decoder outperforms the 3-bit MIM-LUT decoder \cite{meidlinger2020design} by about $0.2$ dB, and also performs better than the 3-bit non-uniform QBP decoder \cite{	Lee05}.
Moreover, the 4-bit MIM-QBP decoder shows almost the same performance as the 4-bit non-uniform decoder by using 2 less bits for the CN update.
In addition, the 4-bit MIM-QBP decoder achieves better performance than the 4-bit MIM-LUT decoder \cite{meidlinger2020design}, and approaches the performance of the floating-point BP decoder within $0.05$ dB at the BER of $10^{-6}$.

\end{example}

\section{Conclusion}\label{section: conclusion}

In this paper, we have proposed the MIM-QBP decoding for regular LDPC codes, which follows the RCQ decoding architecture.
Notably, we have established the first complete and systematic design framework for the RCQ parameters with respect to that, setting proper $(|\mc{L}|, |\mc{R}|, |\mc{S}|, I_\text{max}, q_c, q_v, \sigma_d)$ is sufficient for the design of the optimal/near-optimal RCQ parameters which are able to maximize $I(X; S)$ and $I(X; R)$ for sufficiently large $q_c$ and $q_v$, respectively.
In terms of error rate performance, simulation results showed that the MIM-QBP decoder can always considerably outperform the state-of-the-art MIM-LUT decoder \cite{ Romero16,  Lewandowsky18,  meidlinger2020design}.
Moreover, the MIM-QBP decoder has advantages over the floating-point BP decoder when
\begin{itemize}
    \item the maximum allowed number of decoding iterations is small (generally less than 30), and/or
    \item the code rate is high, and/or
    \item the operating SNR is high.
\end{itemize}
\appendices
\section{Proof of Theorem \ref{theorem: phi_c* is optimal}}\label{appendix: phi_c* is optimal}

Note that $Q_c$ realizes the quantization from $\mc{R}^{d_c - 1}$ to $\mc{S}$ with two steps:
First, convert $\mc{R}^{d_c - 1}$ to $\mc{A}$ based on $\Phi_c$ given by \eqref{eqn: def of Phi_c}, and then quantize $\mc{A}$ to $\mc{S}$ based on the quantizer $\Lambda_c$ given by \eqref{eqn: def of Lambda_c}.
Therefore, to prove Theorem \ref{theorem: phi_c* is optimal} (i.e., $Q_c$ can maximize $I(X; S)$ among all quantizers from $\mc{R}^{d_c - 1}$ to $\mc{S}$), it suffices to prove that the following two statements hold:
\begin{itemize}
	\item   The optimal quantizer from $\mc{A}$ to $\mc{S}$ can achieve the same maximum $I(X; S)$ as the optimal quantizer from $\mc{R}^{d_c - 1}$ to $\mc{S}$.
	\item   $\Lambda_c$ is an optimal quantizer from $\mc{A}$ to $\mc{S}$ which can maximize $I(X; S)$.
\end{itemize}

Let $\phi_c = \phi_c^*$.
For $\mb{r}, \mb{r}' \in \mc{R}^{d_c - 1}$, assume $\Phi_c(\mb{r}) = a_i  \in \mc{A}$ and $\Phi_c(\mb{r'}) = a_{i'}  \in \mc{A}$, where $\Phi_c$ and $\mc{A}$ are given by \eqref{eqn: def of Phi_c} and \eqref{eqn: def of mc_A}, respectively.
We first prove
\begin{equation}\label{eqn: i < i' for phi_c}
\frac{P_{\mb{R}|X}(\mb{r} | 0) }{ P_{\mb{R}|X}(\mb{r} | 1)} > \frac{ P_{\mb{R}|X}(\mb{r}' | 0) }{ P_{\mb{R}|X}(\mb{r}' | 1)} \Rightarrow i < i',
\end{equation}
which is sufficient for proving the above two statements.

\ti{Proof of \eqref{eqn: i < i' for phi_c}:}
We start the proof of \eqref{eqn: i < i' for phi_c} by showing some properties of $\Phi_c$.
For $\mb{r} \in \mc{R}^{d_c - 1}$, we have
\begin{equation*}
g(\mb{r}) = \prod_{i = 1}^{\dim(\mb{r})} g(r_i)
\end{equation*}
according to \eqref{eqn: P(R|0) - P(R|1)}.
Let
\[
h(\mb{r}) = |\{r_i: 1 \leq i \leq \dim(\mb{r}), |g(r_i)| = 1\}|.
\]
Then, we have
\begin{align}
\label{eqn: |Phi_c| = g(r) h(r)epsilon}
|\Phi_c(\mb{r})| &= \sum_{i = 1}^{\dim(\mb{r})} |\phi^*_c(r_i)|\nonumber\\[-0.3em]
&= -  \sum_{i = 1}^{\dim(\mb{r})} \log \left(|g(r_i)| \right) + h(\mb{r}) \epsilon\nonumber\\[-0.3em]
&= -  \log \left(\prod_{i = 1}^{\dim(\mb{r})} |g(r_i)| \right)  + h(\mb{r}) \epsilon\nonumber\\[-0.3em]
&= -  \log \left(|g(\mb{r})| \right) + h(\mb{r}) \epsilon.
\end{align}
Meanwhile, we have
\begin{align}\label{eqn: sgn of Phi_c}
\mr{sgn}(\Phi_c(\mb{r})) &= \prod_{i = 1}^{\dim(\mb{r})} \mr{sgn}(\phi_c(r_i))\nonumber\\[-0.3em]
&= \prod_{i = 1}^{\dim(\mb{r})} \mr{sgn}(g(r_i)) = \mr{sgn}\left( g(\mb{r}) \right).
\end{align}

We continue to prove \eqref{eqn: i < i' for phi_c}.
We have
\begin{align*}
&P_{\mb{R}|X}(\mb{r} | 0) / P_{\mb{R}|X}(\mb{r} | 1) > P_{\mb{R}|X}(\mb{r}' | 0) / P_{\mb{R}|X}(\mb{r}' | 1)\nonumber\\[-0.3em]
\Rightarrow & P_{X|\mb{R}}(0 | \mb{r}) / P_{X|\mb{R}}(1 | \mb{r}) > P_{X|\mb{R}}(0 | \mb{r}') / P_{X|\mb{R}}(1 | \mb{r}')\nonumber\\[-0.3em]
\Rightarrow & g(\mb{r}) > g(\mb{r}')\nonumber
\end{align*}
If $\mr{sgn}(g(\mb{r})) \neq \mr{sgn}(g(\mb{r}'))$, we have
\begin{align*}
&P_{\mb{R}|X}(\mb{r} | 0) / P_{\mb{R}|X}(\mb{r} | 1) > P_{\mb{R}|X}(\mb{r}' | 0) / P_{\mb{R}|X}(\mb{r}' | 1)\nonumber\\[-0.3em]
\Rightarrow& \mr{sgn}(g(\mb{r})) > \mr{sgn}(g(\mb{r}'))\nonumber\\[-0.3em]
\overset{(a)}{\Rightarrow}& \mr{sgn}(\Phi_c(\mb{r})) \succ \mr{sgn}(\Phi_c(\mb{r}'))\nonumber\\[-0.3em]
{\Rightarrow}& \Phi_c(\mb{r}) \succ \Phi_c(\mb{r}')\nonumber\\[-0.3em]
\overset{(b)}{\Rightarrow}& i < i',
\end{align*}
where $(a)$ and $(b)$ are based on \eqref{eqn: sgn of Phi_c} and \eqref{eqn: order of A}.
Otherwise,  we have $\mr{sgn}(g(\mb{r})) = \mr{sgn}(g(\mb{r}')) \neq 0$, leading to
\begin{align*}
&P_{\mb{R}|X}(\mb{r} | 0) / P_{\mb{R}|X}(\mb{r} | 1) > P_{\mb{R}|X}(\mb{r}' | 0) / P_{\mb{R}|X}(\mb{r}' | 1)\\[-0.3em]
\Rightarrow& \mr{sgn}(g(\mb{r}))|g(\mb{r})| > \mr{sgn}(g(\mb{r}))|g(\mb{r}')|\\[-0.3em]
\Rightarrow& -\mr{sgn}(g(\mb{r}))\log(|g(\mb{r})|) < -\mr{sgn}(g(\mb{r}))\log(|g(\mb{r}')|)\\[-0.3em]
\overset{(c)}{\Rightarrow}& \mr{sgn}(g(\mb{r})) (-\log(|g(\mb{r})|) + h(\mb{r}) \epsilon) < \\[-0.3em]
&\quad\quad\quad\quad\quad\quad  \mr{sgn}(g(\mb{r})) (-\log(|g(\mb{r}')|) + h(\mb{r}') \epsilon)\\[-0.3em]
\overset{(d)}{\Rightarrow}& \Phi_c(\mb{r}) < \Phi_c(\mb{r}')\\[-0.3em]
\Rightarrow& \Phi_c(\mb{r}) \succ \Phi_c(\mb{r}')\\[-0.3em]
\overset{(e)}{\Rightarrow}& i < i',
\end{align*}
where $(c), (d)$ and $(e)$ are based on \eqref{eqn: epsilon}, \eqref{eqn: |Phi_c| = g(r) h(r)epsilon}, and  \eqref{eqn: order of A}, respectively.
At this point, the proof of \eqref{eqn: i < i' for phi_c} is completed.

We are ready to utilize \eqref{eqn: i < i' for phi_c} to prove the aforementioned two statements.

\ti{Proof of the first statement:}
Let $N = |\mc{R}|^{d_c - 1}$.
We list the elements of $\mc{R}^{d_c - 1}$ as $\{\mb{r}_1, \mb{r}_2, \ldots, \mb{r}_N\}$ such that for any $1 \leq j < j' \leq N$, assuming $\Phi_c(\mb{r}_j) = a_i \in \mc{A}$ and $\Phi_c(\mb{r}_{j'}) = a_{i'}  \in \mc{A}$, it holds that $i \leq i'$.
Based on \eqref{eqn: i < i' for phi_c}, we have
\begin{equation}\label{eqn: PR|X decrease}
\frac{P_{\mb{R}|X}(\mb{r}_1 | 0) }{ P_{\mb{R}|X}(\mb{r}_1 | 1)} \geq \frac{P_{\mb{R}|X}(\mb{r}_2 | 0) }{ P_{\mb{R}|X}(\mb{r}_2 | 1)} \geq \cdots \geq \frac{P_{\mb{R}|X}(\mb{r}_N | 0) }{ P_{\mb{R}|X}(\mb{r}_N | 1)},
\end{equation}
and for any $\Phi_c(\mb{r}_j) = \Phi_c(\mb{r}_{j'}) = a_i \in \mc{A}$, we have
\begin{equation}\label{eqn: PR|X merge}
\frac{P_{\mb{R}|X}(\mb{r}_j | 0) }{ P_{\mb{R}|X}(\mb{r}_j | 1)}= \frac{P_{\mb{R}|X}(\mb{r}_{j'} | 0) }{ P_{\mb{R}|X}(\mb{r}_{j'} | 1)}= \frac{P_{A|X}(a_i | 0) }{ P_{A|X}(a_i | 1)}.
\end{equation}
The above \eqref{eqn: PR|X merge} indicates that some elements in $\mc{R}^{d_c - 1}$ with the same LLR value are merged into  a single element in $\mc{A}$.
However, the optimal quantizer from $\mc{A}$ to $\mc{S}$ can achieve the same maximum $I(X; S)$ as the optimal quantizer from $\mc{R}^{d_c - 1}$ to $\mc{S}$ \cite{Kurkoski14}.

\ti{Proof of the second statement:}
On the other hand, with \eqref{eqn: i < i' for phi_c} and \eqref{eqn: PR|X decrease}, we have
\begin{equation}
\frac{P_{A|X}(a_1 | 0) }{ P_{A|X}(a_1 | 1)} \geq \frac{P_{A|X}(a_2 | 0) }{ P_{A|X}(a_2 | 1)} \geq \cdots \geq \frac{P_{A|X}(a_{|\mc{A}|} | 0) }{ P_{A|X}(a_{|\mc{A}|} | 1)}.
\end{equation}
As a result, the optimal SDQ $\Lambda_c: \mc{A} \to \mc{S}$ given by \eqref{eqn: def of Lambda_c} can maximize $I(X; S)$ among all quantizers from $\mc{A}$ to $\mc{S}$, as explained in Section \ref{section: DP quantization}.
The proof of Theorem \ref{theorem: phi_c* is optimal} is completed.

\section{Proof of Theorem \ref{theorem: phi_v* is optimal}}\label{appendix: phi_v* is optimal}

Note that $Q_v$ realizes the quantization from $\mc{L} \times \mc{S}^{d_v - 1}$ to $\mc{R}$ with two steps:
First, convert $\mc{L} \times \mc{S}^{d_v - 1}$ to $\mc{B}$ based on $\Phi_v$ given by \eqref{eqn: def of Phi_v}, and then quantize  $\mc{B}$ to $\mc{R}$ based on $\Lambda_v$ given by \eqref{eqn: def of Lambda_v}.
Therefore, to prove Theorem \ref{theorem: phi_v* is optimal} (i.e., $Q_v$ can maximize $I(X; R)$ among all quantizers from $\mc{L} \times \mc{S}^{d_v - 1}$ to $\mc{R}$), it suffices to prove that the following two statements hold:
\begin{itemize}
	\item   The optimal quantizer from $\mc{B}$ to $\mc{R}$ can achieve the same maximum $I(X; R)$ as the optimal quantizer from $\mc{L} \times \mc{S}^{d_v - 1}$ to $\mc{R}$.
	\item   $\Lambda_v$ is an optimal quantizer from $\mc{B}$ to $\mc{R}$ which can maximize $I(X; R)$.
\end{itemize}

Let $\phi_v = \phi_v^*$ and $\phi_{ch} = \phi_{ch}^*$.
For $(l, \mb{s}), (l', \mb{s}') \in \mc{L} \times \mc{S}^{d_v - 1}$, assume $\Phi_v(l, \mb{s}) = b_i  \in \mc{B}$ and $\Phi_v(l', \mb{s}') = b_{i'}  \in \mc{B}$, where $\Phi_c$ and $\mc{A}$ are given by \eqref{eqn: def of Phi_v} and \eqref{eqn: def of mc_B}, respectively.
We first prove
\begin{align}\label{eqn: i < i' for phi_v}
\frac{P_{L, \mb{S} | X}(l, \mb{s} | 0) }{ P_{L, \mb{S} | X}(l, \mb{s} | 1)} >
\frac{ P_{L, \mb{S} | X}(l', \mb{s}' | 0) }{ P_{L, \mb{S} | X}(l', \mb{s}' | 1)}
\Rightarrow  i < i'.
\end{align}
which is sufficient for proving the above two statements.
%

\ti{Proof of \eqref{eqn: i < i' for phi_v}:}
For $(l, \mb{s}) \in \mc{L} \times \mc{S}^{d_v - 1}$, according to \eqref{eqn: def of Phi_v}, we have
\begin{align}\label{eqn: Phi_v = log prod}
\Phi_v(l, \mb{s}) = \log\left(\frac{P_{L|X}(l|0)}{ P_{L|X}(l|1)} \prod_{i=1}^{\dim(\mb{s})} \frac{P_{S|X}(s_i|0)}{ P_{S|X}(s_i|1)}\right).
\end{align}
Then, we have
\begin{align*}
&\frac{P_{L, \mb{S} | X}(l, \mb{s} | 0) }{ P_{L, \mb{S} | X}(l, \mb{s} | 1)} >
\frac{ P_{L, \mb{S} | X}(l', \mb{s}' | 0) }{ P_{L, \mb{S} | X}(l', \mb{s}' | 1)}\nonumber\\
\overset{(f)}{\Rightarrow} & \frac{P_{L|X}(l|0)}{ P_{L|X}(l|1)} \prod_{i=1}^{\dim(\mb{s})} \frac{P_{S|X}(s_i|0)}{ P_{S|X}(s_i|1)} > \nonumber\\[-0.3em]
&\quad\quad\quad\quad   \frac{P_{L|X}(l'|0)}{ P_{L|X}(l'|1)} \prod_{i=1}^{\dim(\mb{s}')} \frac{P_{S|X}(s'_i|0)}{ P_{S|X}(s'_i|1)}\nonumber\\[-0.3em]
\overset{(g)}{\Rightarrow} & \Phi_v(l, \mb{s}) > \Phi_v(l', \mb{s}')\nonumber\\[-0.3em]
\overset{(h)}{\Rightarrow} & i < i',
\end{align*}
where $(f), (g)$ and $(h)$ hold because of \eqref{eqn: joint P_L,S|X}, \eqref{eqn: Phi_v = log prod}, and \eqref{eqn: order of B}, respectively.
At this point, the proof of \eqref{eqn: i < i' for phi_v} is completed.

We are ready to utilize \eqref{eqn: i < i' for phi_v} to prove the aforementioned two statements.

\ti{Proof of the first statement:}
Let $N = |\mc{L}| \cdot |\mc{S}|^{d_v - 1}$.
We list the elements of $\mc{L} \times \mc{S}^{d_v - 1}$ as $\{\mb{y}_1, \mb{y}_2, \ldots, \mb{y}_N\}$ such that for any $1 \leq j < j' \leq N$, assuming $\Phi_v(\mb{y}_j) = b_i \in \mc{B}$ and $\Phi_v(\mb{y}_{j'}) = b_{i'}  \in \mc{B}$, it holds that $i \leq i'$.
Based on \eqref{eqn: i < i' for phi_v}, we have
\begin{equation}\label{eqn: PLS|X decrease}
\frac{P_{L, \mb{S}|X}(\mb{y}_1 | 0) }{ P_{L, \mb{S}|X}(\mb{y}_1 | 1)} \geq \frac{P_{L, \mb{S}|X}(\mb{y}_2 | 0) }{ P_{L, \mb{S}|X}(\mb{y}_2 | 1)} \geq \cdots \geq \frac{P_{L, \mb{S}|X}(\mb{y}_N | 0) }{ P_{L, \mb{S}|X}(\mb{y}_N | 1)},
\end{equation}
and for any $\Phi_v(\mb{y}_j) = \Phi_v(\mb{y}_{j'}) = b_i \in \mc{B}$, we have
\begin{equation}\label{eqn: PLS|X merge}
\frac{P_{L, \mb{S}|X}(\mb{y}_j | 0) }{ P_{L, \mb{S}|X}(\mb{y}_j | 1)}= \frac{P_{L, \mb{S}|X}(\mb{y}_{j'} | 0) }{ P_{L, \mb{S}|X}(\mb{y}_{j'} | 1)}= \frac{P_{B|X}(b_i | 0) }{ P_{B|X}(b_i | 1)}.
\end{equation}
The above \eqref{eqn: PLS|X merge} indicates that some elements in $\mc{L} \times \mc{S}^{d_v - 1}$ with the same LLR value are merged into a single element in $\mc{B}$.
However, the optimal quantizer from $\mc{B}$ to $\mc{R}$ can achieve the same maximum $I(X; R)$ as the optimal quantizer from $\mc{L} \times \mc{S}^{d_v - 1}$ to $\mc{R}$ \cite{Kurkoski14}.

\ti{Proof of the second statement:}
On the other hand, with \eqref{eqn: i < i' for phi_v} and \eqref{eqn: PLS|X decrease}, we have
\begin{equation}
\frac{P_{B|X}(b_1 | 0) }{ P_{B|X}(b_1 | 1)} \geq \frac{P_{B|X}(b_2 | 0) }{ P_{B|X}(b_2 | 1)} \geq \cdots \geq \frac{P_{B|X}(b_{|\mc{B}|} | 0) }{ P_{B|X}(b_{|\mc{B}|} | 1)}.
\end{equation}
As a result, the optimal SDQ $\Lambda_v: \mc{B} \to \mc{R}$ given by \eqref{eqn: def of Lambda_v} can maximize $I(X; R)$ among all quantizers from $\mc{B}$ to $\mc{R}$, as explained in Section \ref{section: DP quantization}.
The proof of Theorem \ref{theorem: phi_v* is optimal} is completed.

\section*{Acknowledgment}

This work was supported by National Natural Science Foundation of China (Grant Nos. 62101462 and 62201258), by Natural Science Foundation of Sichuan (Grant No. 2022NSFSC0952), by Singapore Ministry of Education Academic Research Fund Tier 2 T2EP50221-0036 and MOE2019-T2-2-123.

\ifCLASSOPTIONcaptionsoff
  \newpage
\fi

\bibliographystyle{IEEEtran}
\bibliography{myreference}

\begin{thebibliography}{10}
\providecommand{\url}[1]{#1}
\csname url@samestyle\endcsname
\providecommand{\newblock}{\relax}
\providecommand{\bibinfo}[2]{#2}
\providecommand{\BIBentrySTDinterwordspacing}{\spaceskip=0pt\relax}
\providecommand{\BIBentryALTinterwordstretchfactor}{4}
\providecommand{\BIBentryALTinterwordspacing}{\spaceskip=\fontdimen2\font plus
\BIBentryALTinterwordstretchfactor\fontdimen3\font minus
  \fontdimen4\font\relax}
\providecommand{\BIBforeignlanguage}[2]{{%
\expandafter\ifx\csname l@#1\endcsname\relax
\typeout{** WARNING: IEEEtran.bst: No hyphenation pattern has been}%
\typeout{** loaded for the language `#1'. Using the pattern for}%
\typeout{** the default language instead.}%
\else
\language=\csname l@#1\endcsname
\fi
#2}}
\providecommand{\BIBdecl}{\relax}
\BIBdecl

\bibitem{Gallager62}
R.~G. Gallager, ``Low-density parity-check codes,'' \emph{IRE Trans. Inf.
  Theory}, vol.~8, no.~1, pp. 21--28, Jan. 1962.

\bibitem{chen2018rate}
P.~Chen, K.~Cai, and S.~Zheng, ``Rate-adaptive protograph {LDPC} codes for
  multi-level-cell {NAND} flash memory,'' \emph{IEEE Commun. Lett.}, vol.~22,
  no.~6, pp. 1112--1115, Jun. 2018.

\bibitem{aslam2017edge}
C.~A. Aslam, Y.~L. Guan, and K.~Cai, ``Edge-based dynamic scheduling for
  belief-propagation decoding of {LDPC} and {RS} codes,'' \emph{IEEE Trans.
  Commun.}, vol.~65, no.~2, pp. 525--535, Feb. 2017.

\bibitem{Chen05}
J.~Chen, A.~Dholakia, E.~Eleftheriou, M.~P. Fossorier, and X.-Y. Hu,
  ``Reduced-complexity decoding of {LDPC} codes,'' \emph{IEEE Trans. Commun.},
  vol.~53, no.~8, pp. 1288--1299, Aug. 2005.

\bibitem{Zhang2009qbp}
Z.~Zhang, L.~Dolecek, B.~Nikolic, V.~Anantharam, and M.~J. Wainwright, ``Design
  of {LDPC} decoders for improved low error rate performance: quantization and
  algorithm choices,'' \emph{IEEE Trans. Commun.}, vol.~57, no.~11, pp.
  3258--3268, Nov. 2009.

\bibitem{Richardson01capacity}
T.~J. Richardson and R.~L. Urbanke, ``The capacity of low-density parity-check
  codes under message-passing decoding,'' \emph{IEEE Trans. Inf. Theory},
  vol.~47, no.~2, pp. 599--618, Feb. 2001.

\bibitem{Lee05}
J.-S. Lee and J.~Thorpe, ``Memory-efficient decoding of {LDPC} codes,'' in
  \emph{Proc. IEEE Int. Symp. Inf. Theory}, Sep. 2005, pp. 459--463.

\bibitem{Thorpe02}
\BIBentryALTinterwordspacing
J.~Thorpe, ``Low-complexity approximations to belief propagation for {LDPC}
  codes,'' Oct. 2002. [Online]. Available:
  \url{http://www.systems.caltech.edu/~jeremy/research/papers/low-complexity.pdf}
\BIBentrySTDinterwordspacing

\bibitem{Romero16}
F.~J.~C. Romero and B.~M. Kurkoski, ``{LDPC} decoding mappings that maximize
  mutual information,'' \emph{IEEE J. Sel. Areas Commun.}, vol.~34, no.~9, pp.
  2391--2401, Sep. 2016.

\bibitem{Lewandowsky18}
J.~Lewandowsky and G.~Bauch, ``Information-optimum {LDPC} decoders based on the
  information bottleneck method,'' \emph{IEEE Access}, vol.~6, pp. 4054--4071,
  Jan. 2018.

\bibitem{meidlinger2020design}
M.~Meidlinger, G.~Matz, and A.~Burg, ``Design and decoding of irregular {LDPC}
  codes based on discrete message passing,'' \emph{IEEE Trans. Commun.},
  vol.~68, no.~3, pp. 1329--1343, Mar. 2020.

\bibitem{MacKay1999SPA}
D.~J.~C. MacKay, ``Good error-correcting codes based on very sparse matrices,''
  \emph{IEEE Trans. Inf. Theory}, vol.~45, no.~2, pp. 399--431, Mar. 1999.

\bibitem{he2019onmutual}
X.~He, K.~Cai, and Z.~Mei, ``On mutual information-maximizing quantized belief
  propagation decoding of {LDPC} codes,'' in \emph{Proc. IEEE Global Commun.
  Conf.}, Dec. 2019, pp. 1--6.

\bibitem{Wang2022rcq}
L.~Wang, C.~Terrill, M.~Stark, Z.~Li, S.~Chen, C.~Hulse, C.~Kuo, R.~D. Wesel,
  G.~Bauch, and R.~Pitchumani, ``Reconstruction-computation-quantization
  {(RCQ)}: A paradigm for low bit width {LDPC} decoding,'' \emph{IEEE Trans.
  Commun.}, vol.~70, no.~4, pp. 2213--2226, Apr. 2022.

\bibitem{Mohr2021iblayer}
P.~Mohr, G.~Bauch, F.~Yu, and M.~Li, ``Coarsely quantized layered decoding
  using the information bottleneck method,'' in \emph{Proc. IEEE Int. Conf.
  Commun.}, Jun. 2021, pp. 1--6.

\bibitem{Kang2022qms}
P.~Kang, K.~Cai, X.~He, and J.~Yuan, ``Memory efficient mutual
  information-maximizing quantized min-sum decoding for rate-compatible {LDPC}
  codes,'' \emph{IEEE Commun. Lett.}, vol.~26, no.~4, pp. 733--737, Apr. 2022.

\bibitem{kang2022generalized}
P.~Kang, K.~Cai, X.~He, S.~Li, and J.~Yuan, ``Generalized mutual
  information-maximizing quantized decoding of {LDPC} codes with layered
  scheduling,'' \emph{IEEE Trans. Veh. Tech.}, vol.~71, no.~7, pp. 7258--7273,
  Jul. 2022.

\bibitem{Kang2022qsms}
P.~Kang, K.~Cai, and X.~He, ``Design of mutual-information-maximizing quantized
  shuffled min-sum decoder for rate-compatible quasi-cyclic {LDPC} codes,''
  \emph{Electronics}, vol.~11, no.~19, Oct. 2022.

\bibitem{lv2022qlms}
C.~Lv, X.~He, P.~Kang, K.~Cai, J.~Xing, and X.~Tang, ``Mutual
  information-maximizing quantized layered min-sum decoding of {QC-LDPC}
  codes,'' in \emph{Proc. IEEE Global Commun. Conf.}, Dec. 2022, pp. 1--6.

\bibitem{Kurkoski14}
B.~M. Kurkoski and H.~Yagi, ``Quantization of binary-input discrete memoryless
  channels,'' \emph{IEEE Trans. Inf. Theory}, vol.~60, no.~8, pp. 4544--4552,
  Aug. 2014.

\bibitem{he2021dynamic}
X.~He, K.~Cai, W.~Song, and Z.~Mei, ``Dynamic programming for sequential
  deterministic quantization of discrete memoryless channel,'' \emph{IEEE
  Trans. Commun.}, vol.~69, no.~6, pp. 3638--–3651, Jun. 2021.

\bibitem{IntroAlgo01}
T.~H. Cormen, C.~E. Leiserson, R.~L. Rivest, and C.~Stein, \emph{Introduction
  to Algorithms: 2nd Edition}.\hskip 1em plus 0.5em minus 0.4em\relax
  Cambridge, MA, USA: MIT Press, 2001.

\bibitem{IEEESTD06}
\emph{IEEE standard for information technology---telecommunications and
  information exchange between systems---local and metropolitan area
  networks-specific requirements part 3: {Carrier} sense multiple access with
  collision detection {(CSMA/CD)} access method and physical layer
  specifications}, IEEE Std. 802.3an, Sep. 2006.

\bibitem{ryan2009channel}
W.~Ryan and S.~Lin, \emph{Channel codes: classical and modern}.\hskip 1em plus
  0.5em minus 0.4em\relax Cambridge University Press, 2009.

\bibitem{MacKay}
\BIBentryALTinterwordspacing
D.~J.~C. MacKay, ``Encyclopedia of sparse graph codes.'' [Online]. Available:
  \url{http://www.inference.phy.cam.ac.uk/mackay/codes/data.html}
\BIBentrySTDinterwordspacing

\end{thebibliography}

\end{document}